\documentclass[12pt]{article}
\usepackage{amsmath}
\usepackage{enumerate}
\usepackage{natbib}
\usepackage{setspace}
\renewcommand{\baselinestretch}{1.5} 
\usepackage{amsmath}
\usepackage{enumerate}
\usepackage{bbm}
\usepackage{pgf,tikz}
\usetikzlibrary{decorations.pathreplacing,angles,quotes}
\usepackage{mathrsfs}

\newtheorem{theorem}{Theorem}
\newtheorem{assumption}{Assumption}

\newtheorem{remark}{Remark}
\usepackage{color} 
\usepackage{enumitem} 
\usepackage{mathtools}
\newcommand{\blind}{1}

\usepackage{pifont}
\usepackage{amssymb}
\usepackage{booktabs}
\usepackage{graphicx}
\usepackage{tabularx}
\usepackage{hyperref}
\usepackage{titlesec}

\titleformat{\paragraph}
{\normalfont\normalsize\bOseries}{\theparagraph}{1em}{}
\titlespacing*{\paragraph}
{0pt}{3.25ex plus 1ex minus .2ex}{1.5ex plus .2ex}

\usepackage{apptools}
\AtAppendix{\counterwithin{lemma1}{section}}

\newtheorem{lemma1}{Lemma}[section]
\newtheorem{theorem1}{Theorem}[section]
\usepackage[]{appendix}

\usepackage{GrandMacros}

\newcommand\independent{\protect\mathpalette{\protect\independenT}{\perp}}
\def\independenT#1#2{\mathrel{\rlap{$#1#2$}\mkern2mu{#1#2}}}

\allowdisplaybreaks

\begin{document}

\def\spacingset#1{\renewcommand{\baselinestretch}%
{#1}\small\normalsize} \spacingset{1}


\if1\blind
{
  \title{\large\bf Continuous-time structural failure time model for intermittent treatment}
  \author{\normalsize Guanbo Wang\thanks{CausaLab, Department of Epidemiology, Harvard T.H. Chan School of Public Health, MA 02115, U.S.A. Email: gwang@hsph.harvard.edu}  \thanks{We thank Drs. Mireille E. Schnitzer and Robert W. Platt for the useful suggestions and discussion.},
    Siyi Liu\thanks{Co-first author, Department of Statistics, North Carolina State University, NC 27695, U.S.A.},
     Shu Yang\thanks{Department of Statistics, North Carolina State University, NC 27695, U.S.A. }}
     \date{}
  \maketitle
} \fi

\if0\blind
{
  \bigskip
  \bigskip
  \bigskip
  \begin{center}
    {\bf Title}
\end{center}
  \medskip
} \fi

\bigskip
\begin{abstract}
The intermittent intake of treatment is commonly seen in patients with chronic disease. For example, patients with atrial fibrillation may need to discontinue the oral anticoagulants when they experience a certain surgery and re-initiate the treatment after the surgery. As another example, patients may skip a few days before they refill a treatment as planned. This treatment dispensation information (i.e., the time at which a patient initiates and refills a treatment) is recorded in the electronic healthcare records or claims database, and each patient has a different treatment dispensation. Current methods to estimate the effects of such treatments censor the patients who re-initiate the treatment, which results in information loss or biased estimation. In this work, we present methods to estimate the effect of treatments on failure time outcomes by taking all the treatment dispensation information. The developed methods are based on the continuous-time structural failure time model, where the dependent censoring is tackled by inverse probability of censoring weighting. The estimators are doubly robust and locally efficient. 
\end{abstract}

\noindent%
\vfill

\newpage
\spacingset{1.45} 
\section{Introduction}
When analyzing longitudinal observational data, adjusting for time-dependent confounding to arrive at an unbiased estimation of a causal effect is necessary.  \citep{daniel2013methods,robins1992estimation,robins2000marginal} When the outcome is time-to-event, three categories of models would be considered to deal with time-dependent confounding, namely G-formula \citep{keil2014parametric}, marginal structural Cox models \citep{hernan2000marginal} and structural nested accelerated/cumulative failure time models \citep{picciotto2012structural,robins1992estimation}. The first two categories of models focus on estimating the marginal counterfactual hazard contrast under different treatment regimens, while the latter one focuses on modeling the distribution of the counterfactual failure time or the ratio of mean observed failure time and counterfactual failure time. Unlike G-formula and marginal structural Cox models, structural failure time models can handle time-dependent effect modifiers \citep{robins2000marginal}. In addition, these models relax the positivity assumption required by marginal structure Cox models \citep{robins2014structural}. Despite these advantages, the derivation, estimation, and inference of structural failure time models are quite involved and thus draw less attention. Furthermore, artificial censoring \citep{joffe2001administrative,joffe2012g,rotnitzky1995semiparametric} has been a drawback of structural failure time models.\\
The data collected from Electronic Healthcare Records (EHR) capture longitudinal information on patients' diagnostics, medication prescriptions, comorbidities, etc., over time. Importantly, the information collected, especially the treatment dispensation, may be recorded at any time rather than at regularly spaced intervals. When employing the above methods to analyze such data, researchers are required to pre-process the data to artificially create data measured at regular intervals -- i.e., assuming the values of variables can change only at pre-specified time points for all patients. However, the use of pre-processed data may induce a violation of the sequential randomization assumption and is subject to information loss \citep{andreano2017adherence,robins2014structural,yang2020semiparametric,ferreira2020impact}.\\
Some work has been exploring to estimate treatment effects in a continuous-time setting. Robins (1998) \citep{robins1998structural} conjectured that g-estimation in structural nested accelerated/cumulative failure time models could be extended to the settings with continuous-time processes and rank-preserving assumption holds. Lok (2017) \citep{lok2017mimicking} proved the conjecture without rank preservation. Some work has also explored how to accommodate continuous-time processes in marginal structural models \citep{hu2021estimating,johnson2022treatment,roysland2011martingale,yang2018modeling}. Notably, Yang et al. (2020) \citep{yang2020semiparametric} developed a continuous-time structural failure time model (ctSFTM), which allows for the values of variables to change irregularly space in the setting where treatment dispensation is discontinued permanently. The method posits a proportional hazard model for the treatment discontinuity to take the treatment process into account.\\
However, the method can only accommodate permanent treatment discontinuation, i.e., when a subject resumes the treatment of interest, the method censors the subject. Oftentimes, patients may start and stop the treatment repeatedly over time. That is, temporary treatment discontinuation may be a common phenomenon for treating some diseases. For example, patients would resume the treatment after discontinuation, where the discontinuation may be due to the intake of the treatment's contraindication because of complications or new-developed diseases or undertaking surgery beyond the case where patients fail to adhere to their prescriptions. Such phenomena would be more pronounced when the treatment of interest is for chronic diseases, such as heart or kidney diseases \citep{navaneethan2011development,perreault2020oral}. When the data to be analyzed are collected from EHR, the time of treatment initiation, treatment refill, and treatment dispensation will be available. Censoring the subjects who discontinue the treatment temporarily may cause a great deal of information loss. \\ 
In this work, based on ctSFTM, we develop a method to estimate a treatment effect when the treatment may be taken intermittently for a continuous time. Differing from ctSFTM, we capture the entire treatment trajectory by modeling the hazards of treatment refill after the end of each course of treatment rather than only treatment discontinuity. Treatment dispensations represent when patients pick up their medication, which is related to (but not the same as) the treatment taken by patients. When the treatment dispensation is a proxy to, or can be regarded as a surrogate of actual treatment intake, our method provides a way to alleviate the estimation bias due to non-compliance.\\
The proposed estimator is doubly robust and locally efficient. With the same spirit of ctSFTM, our method also allows for investigating the time-dependent effect modification, which advanced personalized medicine in survival analysis. In addition, rank-preservation is not assumed in the ctSFTM, which further relaxes deterministic relationships of the observed failure time and the counterfactual baseline failure time. Furthermore, artificial censoring is a common practice for addressing censoring in structural nested model, which creates heavy computational burden and non-trivial inference. \citep{joffe2001administrative,joffe2012g} More importantly, the use of artificial censoring would be unrealistic for EHR data, because it requires the assumption of administrative censoring. \citep{rotnitzky1995semiparametric} We tackle those issues by employing inverse probability censoring weighting, where the weighting scheme can be readily used for inference.
\section{Notation, model, and assumptions}
\subsection{Revisit of ctSFTM}
Next, we briefly introduce the ctSFTMs, as developed by Yang et al (2020) \citep{yang2020semiparametric}. Let $\tau_{i}$ be the observed failure time, $A_{iu}$ and $\bL_{iu}$ be binary treatment and $(p-1)$-dimensional covariate processes, respectively. Let $U_{i}$ be the counterfactual baseline failure time had the treatment always been withheld. Assume treatment discontinuation is permanent. The model assumes:
\begin{align}
\label{original model}
   U_{i} \sim U_{i}\left(\bpsi^{*}\right)=\int_{0}^{\tau_{i}} \exp \left[\left\{\psi_{1}^{*}+\bpsi_{2}^{* \intercal} g\left(\bL_{iu}\right)\right\} A_{iu}\right] \mathrm{d} u, 
\end{align}
where $\sim$ stands for has the same distribution as, $g(\bL_{iu})$ is some function of the covariates, and $U_{i}\left(\bpsi^{*}\right)$ is so-called mimicking counterfactual baseline failure time \citep{lok2017mimicking}, which is a function of $\bpsi^{*}$. The model essentially assumes the counterfactual baseline failure time has the same distribution as the mimicking counterfactual baseline failure time, which is modeled by $\bpsi^{*}, A_{iu}, \bL_{iu}$ and $\tau_{i}$. If rank-preserving assumption holds, then $U_{i}=U_{i}\left(\bpsi^{*}\right)$.\\
Consider a simplified model $U_{i} \sim U_{i}\left(\bpsi^{*}\right)=\int_{0}^{\tau_{i}} \exp \left(\psi_{1}^{*} A_{iu}\right) \mathrm{d} u$. We can interpret the parameter of interest $\psi^*_1$ in the following way: the survival time under continuous treatment during the follow-up has the same distribution as the survival time under no treatment during the follow-up distribution that is expanded or contracted by the factor $e^{-\psi^*_1}$ \citep{robins2014structural}. We can see that from the following fact: 1) If $A_{iu}=0$ for $0\leqslant u \leqslant \tau_{i}$, then $U_{i}\sim U_{i}\left(\bpsi^{*}\right)=\tau_{i}$; 2) If $A_{iu}=1,$ for $0\leqslant u \leqslant \tau_{i}$, then $U_{i}\sim U_{i}\left(\bpsi^{*}\right)=\tau_{i}e^{\psi^*_1}$, $\tau_{i}= e^{-\psi^*_1} U_{i}\left(\bpsi^{*}\right)$; 3) If $\psi^*_1=0$, then $0\leqslant u \leqslant \tau_{i}$, then $U_{i}\sim U_{i}\left(\bpsi^{*}\right)=\tau_{i}$. Overall, the model describes how the treatment (possibly along with other characteristics of patients) accelerates or decelerates the failure time $\tau_{i}$ compared to counterfactual baseline failure time $U_{i}$. The multiplicative factor $ -\exp \left[\left\{\psi_{1}^{*}  +\bpsi_{2}^{* \intercal} g\left(\bl_{iu}\right)\right\}\right]$ can be interpreted as the heterogeneous effect rate of the treatment, defined as a ratio of failure time and counterfactual baseline failure time when the rank-preserving assumption holds. It takes a positive value when the treatment is beneficial to the patients with covariate values  $\bl_{iu}, u\in [0,\tau_{i}]$.\\
\subsection{The proposed model}
For now, we assume no censoring (censoring is addressed in section \ref{AppendixCensoring}). Suppose the sample is drawn independently from a large population of interest, and the $n$ subjects in the sample are identically distributed. Let $\tau_i,$ $i=1,\dots,n$ be the time to event for patient $i$.
Define a treatment process $A_{iu}$, taking value 1 if the patient $i$ is taking the treatment at time $u$, and 0 otherwise.
If patients take the treatment constantly, patients have to pick up their prescriptions regularly, every month, for example, as instructed to continue the treatment. However, some patients fail to refill the prescription on time, resulting in a potentially biased estimation of the treatment effect if it is not taken into account. We thus define a series of discrete times $0=V_{i,0}<V_{i,1}<\dots<V_{i,K_{i}}$, where $V_{i,0}$ is the beginning of follow-up and $V_{i,k}$ is the time to the $k^{\text{th}}$ time of pickup the prescription. We assume $K_{i}\geqslant 1$ and thus $\tau_{i}>V_{i,K_{i}}$. The time between the two successive refills is $V_{i,k}-V_{i,k-1}$. Suppose the period that one prescription covers is $w$, which is drug-specific. If the dispensing overlaps this period i.e., $V_{i,k}-V_{i,k-1}<w$, then we set $V_{i,k}$ to $V_{i,k-1}+w$ \citep{perreault2020oral}. Therefore, the trajectory of $A_{iu}$ can be described by $V_{i,k}$ and $w$. We define the gaping time $T_{ik}$ as the time elapse from the end of last prescription to the next treatment refill. Thus, $T_{ik}=V_{i,k}-(V_{i,k-1}+w^{-})\geqslant 0$, where $w^{-}=w-\epsilon$ and $\epsilon$ is a time that sufficiently short, for example, $10^{-6}$. We define the gaping time as above so that we can define a valid counting process next. We say $N_{ik}(u)=\mathbbm{1}(u\geqslant T_{ik})$, and $Y_{ik}(u)=\mathbbm{1}(u\leqslant T_{ik})$ are the counting process (of if the $k^{\text{th}}$ non-terminal pickup occurs before or at time $u$) and at-risk process respectively. A Figure that illustrates this is seen in Supplementary material \ref{Figure}.\\
Denote $\bL_{iu}$ as the $(p-1)$-dimensional covariate process. For the purpose of regularity, we assume that all continuous-time processes are continuous from the right, have limits from the left (a.k.a. cadlag processes), and are square integrable. To further simplify the notation, we denote $\bH_{iu}\coloneqq(\bL_{iu},\bA_{iu-})$ as the combined process, which is the treatment dispensation process right before $u$ together with the covariate process at $u$. We also denote  $\overline{\bH}_{iu} \coloneqq (\overline{\bL}_{iu},\overline{\bA}_{iu^{-}})$ as the history process up until $u$, which is a collection of combined processes up until $u$. Thus the observables in the data can be represented as $\bO_{i}=\{\tau_{i}, \overline{\bH_{i}}_{\tau_{i}}\}$. We define our parameter of interest as a $ p$-dimensional vector $\bpsi=\{\psi_1, \bpsi_2\}$, and the true value of the parameters $\bpsi^{*}=\{\psi_1 , \bpsi_2\}$.\\
To facilitate future development, we need to define the potentially counterfactual failure time $U_{i}$ had all the treatments always been withheld during the follow-up, and $U_{i}(\bpsi^{*})$ as a random variable that has the same distribution with $U_{i}$, and can be expressed by the parameter of interest $\bpsi^{*}$.
Our aim is to estimate the effects of treatment dispensation and investigate how other covariates modify the effects. We assume the same model as (\ref{original model}). Note that in the method developed by Yang et al. (2020) \citep{yang2020semiparametric}, there is a restriction on $A_{iu}$: it can only change from 1 to 0, and will be censored if it changes back to 1. The method developed below relaxes this restriction. Before we move forward, two remarks need to be made about the model.
\begin{remark}
\label{constant heterogeneous treatment effect}
\textnormal{\textbf{(Constant Heterogeneous Treatments Effects)}}
The effects of treatment dispensation and how effect modifiers alter the treatment dispensation effect are constant over time. That is, the parameter of interest $\bpsi^{*}$ is not a function of time $u$.
\end{remark}
Remark \ref{constant heterogeneous treatment effect} states that the model assumes the heterogeneous treatment effect is constant over time. When a patient resumes the treatment, the model assumes the effects would remain the same as the effects when the patient first took the treatment. Therefore, the model would be invalid if the treatment or effect modifiers' effect changes with time and the overall health condition of patients.
\begin{remark}
\label{no legacy effect}
\textnormal{\textbf{(No Legacy Effect)}}
The model assumes that when patients discontinue treatment, the effect of the treatment vanishes immediately.
\end{remark}
Essentially, the model assumes that the treatment has no delay or legacy effects. Therefore, the model is valid if the treatment affects the outcome instantly, and its effect disappears at a relatively fast rate. For example, a treatment whose half-life is short would satisfy the assumption, which is the case for OACs.\\
To identify and estimate the parameter of interest, more assumptions are needed.
\begin{assumption}
\label{NUC}
\textnormal{\textbf{(No Unmeasured Confounding)}}
The hazard of the $k^\mathrm{th}$ treatment refill is $\forall u\in[V_{i,k-1}+w,V_{i,k}]$
\begin{align*}
\lambda_{ik}(u | \bO_{i}, U_{i}) &=\lim _{\text{d}u \rightarrow 0} \frac{1}{\text{d}u} P(u\leqslant T_{ik}<u+\text{d}u | \bO_{i}, U_{i}, T_{ik} \geqslant u) \\
&=\lim _{\text{d}u \rightarrow 0} \frac{1}{\text{d}u} P\left(u \leqslant T_{ik}<u+\text{d}u | \overline{\bH}_{i,V_{k-1}+w+u}, T_{ik}\geqslant u\right)
=\lambda_{ik}\left(u | \overline{\bH}_{i,V_{k-1}+w+u}\right).
\end{align*}
\end{assumption}
Assumption \ref{NUC} implies that $\lambda_{ik}(u | \bO_{i}, U_{i})$ depends only on the past treatment dispensation and covariate history until time $V_{k-1}+w+u$, denoted $\overline{\bH}_{i,V_{k-1}+w+u}$, but not on the future variables and $U_{i}$ \citep{yang2020semiparametric}. That is, the relationship between outcome at $u$ and treatment dispensation is only confounded by the past information but not by the future. This assumption holds if all the factors that affect the hazard of pickup at $V_{i,k-1}+w+u$ are being considered in the data collected. 
We also show in Supplementary material \ref{after Assumption 3 appendix} that
\begin{align}
\label{treatment switching hazard equivalence}
\lambda_{ik}\left\{u |U_{i}(\bpsi^{*}), \overline{\bH}_{i,V_{k-1}+w+u}\right\}=\lambda_{ik}\left(u | \overline{\bH}_{i,V_{k-1}+w+u}\right), \forall k.
\end{align}
This further implies that the treatment  dispensation process given history is independent of $U_{i}(\bpsi^{*})$.\\
Denote $\sigma(\bH_{iu}: t\geqslant0)$ as the $\sigma$-field generated by $\bH_u$, and $\sigma(\overline{\bH}_{iu}: t\geqslant0)$ as a $\sigma$-field generated by $\cup_{s\leqslant u}\sigma(\overline{\bH}_{s})$. Therefore, under common regularity conditions for the counting process, we show in Supplementary material \ref{martingale proof} that
\begin{align}
\label{martingale}
M_{ik}(u)=N_{ik}(u)-\int_{0}^{u} \lambda_{ik}(s|\overline{\bH}_{i,V_{k-1}+w+s})Y_{ik}(s)ds
\end{align} 
is a martingale that is generated by the filtration $\mathscr{F}_{i,V_{k-1}+w+u}=\sigma\{U_{i}(\bpsi^{*}),\overline{\bH}_{i,V_{k-1}+w+u}: u\geqslant0\}$, where $\mathscr{F}_{i,V_{k-1}+w+u}$ is an increasing sequence of $\sigma$-fields. This enables us to identify the parameters of interest $\bpsi^{*}$.
\begin{assumption}
\label{independent gapping time}
\textnormal{\textbf{(Conditional Independent Counting Processes)}}
The jumps of multiple counting processes that are defined on the same subject at time $u$ are independent of each other conditional on the filtration generated by the covariates in such time periods. 
\end{assumption}
Mathematically, this assumption states that $\{\Delta N_{i1}(u), \dots, \Delta N_{iK_{i}}(u)\}$ are independent 0-1 random variables given all the filtration $\mathscr{F}_{i,V_{k-1}+w\rightarrow V_{k-1}+w+u}=\sigma\{\overline{\bH}_{i,V_{k-1}+w\rightarrow V_{k-1}+w+u}: u\geqslant0\}, \forall k=1,\dots,K_{i}$, where $\overline{\bH}_{i,V_{k-1}+w\rightarrow V_{k-1}+w+u}$ is $\overline{\bH}$ from $V_{k-1}+w$ to $V_{k-1}+w+u$ for subject $i$. This assumption allows us to identify the covariance of two martingale transforms, as per the Lemma shown in Supplementary material \ref{martingale_lemma}.\\
Here, we take an example to illustrate when this assumption holds. Suppose a prescription lasts for 15 days, and a patient picks up the prescription at day 1, 20 and 40. For any time between days 15-20 and days 35-40, for instance, days 18 and 38, the assumption holds when whether the patient refills the prescription at day 38 (Decision 2) is independent of whether they do so at day 18 (Decision 1) given the values of covariates between days 15-18 and days 35-38. It is not saying that Decision 2 is independent of the covariate history between days 1-38, as assumed in Assumption \ref{NUC}, but assumes it is only independent of Decision 1 if we know the covariates' value between days 15-18 and days 35-38 -- the period when the patient fails to adhere to the prescription because we believe that Decision 2 depends on Decision 1 only through the covariate history in the subset of the patients who do not take the treatment. This is reasonable when we assume that, if a patient skips the treatment, then when the treatment will be refilled is decided by his/her recent health conditions, specifically, the conditions since treatment discontinuation, rather than the conditions before. This assumption is more likely to hold when the heterogeneous treatment effects are constant over time.
\section{Estimation}
\label{estiamtion 4.2.2}
In this section, we provide the semiparametric estimation of $\bpsi^{*}$, i.e, we derive a regular and asymptotically linear estimator $\widehat{\bpsi}$ that satisfies: $
    n^{\frac{1}{2}}(\widehat{\bpsi}-\bpsi^{*})=\mathbb{P}_n\Phi(\bO_{i})+o_P(1),
$
where $\mathbb{P}_n$ is the empirical measure induced by $\bO_{i}$ such that $\mathbb{P}_n\Phi(\bO_{i})=\frac{1}{n}\sum_{i=1}^{n}\Phi(\bO_{i})$, and $\Phi(\bO_{i})$ is an influence function, which is mean zero and has finite and non-singular variance. We use these properties to construct the estimating equation to estimate $\widehat{\bpsi}$. In Theorem \ref{theorem 1}, we give the orthogonal complement of the nuisance tangent space $\Lambda^{\perp}$, to which the influence functions belong. 
\begin{theorem}
\label{theorem 1}
Under Model (\ref{original model}) and Assumption \ref{NUC} and 
\ref{independent gapping time}, denote $h(\cdot)$ is an arbitrary function, the orthogonal complement of the nuisance tangent space for $\bpsi^{*}$ is
\begin{align*}
\begin{split}
    \Lambda^{\perp}=
    &\sum_{k=1}^{K_{i}}\int\Bigg\{ h\left\{u,U_{i}(\bpsi^{*}), \overline{\bH}_{i, V_{k-1}+w+u}\right\}-\Ebb\Big[h\left\{u,U_{i}(\bpsi^{*}), \overline{\bH}_{i, V_{k-1}+w+u}\right\} |\overline{\bH}_{i, V_{k-1}+w+u}, \\
    &T_{ik}\geqslant u\Big]\Bigg\}
    \text{d}M_{ik}(u) : h\left\{u,U_{i}(\bpsi^{*}), \overline{\bH}_{i, V_{k-1}+w+u}\right\}\in\Rbb^{ p},
\end{split}
\end{align*}
\end{theorem}
The proof is given in Supplementary material \ref{theorem 1 appendix}. We use $\int$ to denote $\int_{0}^{\infty}$. Let $\btheta$ and $\btheta^{*}$ be the nuisance parameter and its true value, 
$S_{\psi}(\bO_{i})=\partial \log f_{\bO_{i}}(\tau_{i},\overline{H}_{\tau_{i}};\bpsi,\btheta)/ \partial \bpsi$, 
evaluated at $(\bpsi^{*},\btheta^{*})$, be the score function of $\bpsi^{*}$. Then the efficient score of $\bpsi^{*}$ is $S_{\text{eff}}(\bO_{i})=\prod \{S_{\psi}(\bO_{i})|\Lambda^{\perp}\}$, which is the projection of $S_{\psi}(\bO_{i})$ onto the orthogonal complement of the nuisance tangent space. Following \citep{tsiatis2007semiparametric}, the efficient influence function is $\Ebb\{S_{\text{eff}}(\bO_{i})S_{\text{eff}}(\bO_{i})^{\intercal}\}^{-1} S_{\text{eff}}(\bO_{i})$, and $\text{Var}\{\Phi_{\text{eff}}(\bO_{i})\}$ is $\left[\Ebb\{S_{\text{eff}}(\bO_{i})S_{\text{eff}}(\bO_{i})^{\intercal}\}\right]^{-1}$, which is the semiparametric efficiency bound. However, in general, the analytical form of the score of $\bpsi^{*}$ is intractable. In order to proceed with the estimation, we restrict ourselves to a subspace,
of $\Lambda^{\perp}$ with $h\left\{u,U_{i}(\bpsi^{*}), \overline{\bH}_{i, V_{k-1}+w+u}\right\}=c(\overline{\bH}_{i, V_{k-1}+w+u})U_{i}(\bpsi^{*})$, $\forall c(\overline{\bH}_{i, V_{k-1}+w+u})\in\Rbb^{p}$, thus the estimating function is
\begin{align}
\label{estimating equation}
G(\bpsi ; \bO_{i})=\sum_{k=1}^{K_{1}} \int c(\overline{\bH}_{i, V_{k-1}+w+u})\left[U_{i}(\bpsi)-\Ebb\left\{U_{i}(\bpsi) |\overline{\bH}_{i, V_{k-1}+w+u}, T_{ik}\geqslant u\right\}\right] \text{d}M_{ik}(u).
\end{align}
We can further replace the $\bpsi$ by $\bpsi^{*}$ in $G(\cdot)$ due to Assumption \ref{NUC}: $U_{i}(\bpsi^{*})\independent M_{ik}(u)| (\overline{\bH}_{i, V_{k-1}+w+u}, \\T_{ik}\geqslant u)$, which leads to $\Ebb\{G(\bpsi^{*};\bO_{i})\}=0$. Therefore, the estimator can be obtained by solving
\begin{align}
    \label{replaced estimating equation}
    \Pbb_{n}\{G(\bpsi,\bO_{i})\}=0.
\end{align}
We first show in Supplementary material \ref{identifiability1} that under Assumptions 1-3, $\bpsi^{*}$ is identifiable. Then we further show in Supplementary material \ref{c_opt} that within this class of estimator, we can find an optimal choice of $c(\overline{\bH}_{i, V_{k-1}+w+u})$ such that the resulting estimator achieves the semiparametric efficiency bound. The optimal choice is given by
\begin{align}
\begin{split}
\label{copt}
    c^{\text{opt}}(\overline{\bH}_{i, V_{k-1}+w+u})=&\left\{1-\Delta\int \lambda_{ik}\left(u | \overline{\bH}_{i, V_{k-1}+w+u}\right) \text{d}u\right\}^{-1}
     \left[\text{Var}\{U_{i}(\psi)|\overline{\bH}_{i, V_{k-1}+w+u}, T_{ik}\geqslant u\}\right]^{-1}\\
    &\times\Ebb\Bigg(\frac{\partial}{\partial \bpsi}\left[ U_{i}(\bpsi)-\Ebb\{U_{i}(\bpsi)|\overline{\bH}_{i, V_{k-1}+w+u}, T_{ik}\geqslant u\}\right]|\overline{\bH}_{i, V_{k-1}+w+u}, T_{ik}= u\Bigg).
\end{split}
\end{align}
Replacing $c(\overline{\bH}_{i, V_{k-1}+w+u})$ with $c^{\text{opt}}(\overline{\bH}_{i, V_{k-1}+w+u})$ in (\ref{replaced estimating equation}) will give us the most precise regular asymptotically linear semiparametric estimator, under the correct specification of two models. These models are the counterfactual failure time model: $\Ebb\{U_{i}(\bpsi)|\overline{\bH}_{i, V_{k-1}+w+u}, T_{ik}= u;\xi\}$ and the treatment model: $\lambda_{ik}\left(u | \overline{\bH}_{i,V_{k-1}+w+u};\gamma_{V}\right)$.The resulting estimator is a doubly-robust estimator by Theorem \ref{DR}, with proof in Supplementary material \ref{DR_proof}. Furthermore, the analytical form of the semiparametric efficiency bound is also given in Supplementary material \ref{DR_proof}.
\begin{theorem}
\label{DR}
Under Model (\ref{original model}) and Assumption \ref{NUC}, the estimating equation (\ref{replaced estimating equation}) with the optimal choice of $c(\cdot)$ provided in (\ref{copt}) is unbiased for zero if either the counterfactual failure time model or treatment model is correctly specified, but not necessarily both. In addition, the estimator is semi parametrically locally efficient in the class of estimators where $h\left\{u,U_{i}(\bpsi^{*}), \overline{\bH}_{i, V_{k-1}+w+u}\right\}=c(\overline{\bH}_{i, V_{k-1}+w+u})U_{i}(\bpsi^{*})$. 
\end{theorem}
The next section shows how to handle censoring using IPCW and gives detailed procedures to estimate the parameter of interest in the presence of censoring. 
\section{Censoring}
\label{AppendixCensoring}
We follow the work by \citep{yang2020semiparametric} and adapt its framework on our scenario. IPCW \citep{rotnitzky2009analysis} is a potential method to handle censoring, the main idea is to give the subjects who are not being censored more weights if the subjects are more likely to be censored. Let $C_{i}$ be the time of censoring for patient ${i}$. Thus the observed data become $\bO_{i}=\{X_{i}=\text{min}(\tau_{i}, C_{i})\}, \Delta_{i}=\mathbbm{1}(\tau_{i}\leqslant C_{i}), \overline{\bH}_{i,X_{i}}\}$.\\
First, we made an ignorable censoring mechanism assumption:
\begin{assumption}
\label{Ignorable Censoring Mechanism}
\textbf{(Ignorable Censoring Mechanism)}
The hazard of censoring $\forall u\in (V_{i,K_{i}},X_{i}]$ is
\begin{align*}
    \lambda_{iC}(u|\bO_{i},\tau_{i}>u)
    =&\lim _{\text{d}u \rightarrow 0} \frac{1}{\text{d}u} P(u\leqslant C_{i}<u+\text{d}u | C_{i}\geqslant u, \bO_{i}, \tau_{i} \geqslant u) \\
    =&\lim _{\text{d}u \rightarrow 0} \frac{1}{\text{d}u} P\left(u \leqslant C_{i}<u+\text{d}u |C_{i}\geqslant u, \overline{\bH}_{iu}, \tau_{i}\geqslant u\right)
=\lambda_{iC}\left(u | \overline{\bH}_{iu}, \tau_{i}\geqslant u\right),
\end{align*}
which is denoted as $\lambda_{iC}\left(u | \overline{\bH}_{iu}\right)$ for shorthand.
\end{assumption}
It states that the hazard of censoring depends only on the past treatment and covariate history until time $u$, but not on the future variables and failure time. This assumption holds if the set of historical covariates contains all prognostic factors for the failure time that affect the loss to follow-up at time $u$.  This assumption equivalents to missing at random assumption if we view censoring as a type of missing data.\\
This assumption is similar to Assumption \ref{NUC}. However, note that $\lambda_{ik}(u|\overline{\bH}_{i,V_{k-1}+w+u})$ is defined on  $u\in[V_{i,k-1}+w,V_{i,k}], \forall k=1,\dots,K_{i}$, which is some time periods before $V_{i,K_{i}}$; whereas $\lambda_{iC}\left(u | \overline{\bH}_{iu}\right)$ is defined on $u\in(V_{i,K_{i}},X_{i}]$, which is after $V_{i,K_{i}}$, because we assume there are always $K_{i}\geqslant 1$ times of pick-ups being observed, so censoring would not happen before $V_{i,K_{i}}$. Therefore, $\lambda_{ik}(u|\overline{\bH}_{i,V_{k-1}+w+u})=\lambda_{ik}(u|\overline{\bH}_{i,V_{k-1}+w+u}, C_{i}\geqslant u)$. \\
We define the probability of the patient not being censored before $u$ as $S_{Ci}(u|\overline{\bH}_{u})=\exp\{-\int_{V_{i,K_{i}}}^{u}\lambda_{iC}\left(u | \overline{\bH}_{iu}\right)\text{d}u\}$. We further make a positivity assumption for $S_{Ci}(u|\overline{\bH}_{u})$.
\begin{assumption}
\label{Positivity for Survival of Censoring}
\textbf{(Positivity for Censoring)}
There exists a positive constant $\epsilon$ such that $\forall u\in(V_{i,K_{i}},X_{i}], P\{S_{Ci}(u|\overline{\bH}_{u})\geqslant\epsilon\}=1$.
\end{assumption}
With all the assumptions held, we can thus estimate $\bpsi^{*}$ with the presence of censoring.
\begin{theorem}
\label{Theorem3}
Under Assumption 1-4, the estimator solves the following estimating equation is doubly-robust in the class of estimators when $h\left\{u,U_{i}(\bpsi^{*}), \overline{\bH}_{i, V_{k-1}+w+u}\right\}=c(\overline{\bH}_{i, V_{k-1}+w+u})U_{i}(\bpsi^{*})$ given $\lambda_{iC}\left(u | \overline{\bH}_{iu}\right)$ is correctly specified.
\begin{align}
    \Pbb_{n}\left\{\frac{\Delta_{i}}{S_{Ci}(\tau_{i}|\overline{\bH}_{\tau_{i}})}G(\bpsi,\bO_{i})\right\}=0.
\end{align}
\end{theorem}
The proof is shown in Supplementary material \ref{theorem3}. With a similar augment that we provided in Supplementary material \ref{identifiability1}, $\bpsi^{*}$ is identifiable.\\
Now similar to \citep{yang2020semiparametric}, we can sketch the algorithm to estimate $\bpsi^{*}$\\
\textbf{Step 1.} Using the data $(V_{i,k}, \overline{H}_{i, V_{k}})$ to fit a time-dependent Accelerated Failure Time model \citep{aftreg}, or a nonparametric model \citep{lee2017boosted} (without censoring, regarding $V_{i,k}$ as time to event): to obtain the estimate of $\lambda_{k}(u|\overline{\bH}_{iu})$: $\widehat{\lambda}_{k}(u|\overline{\bH}_{iu})$. Then, we can obtain the estimated martingales 
\begin{align*}
    \widehat{M}_{ik}(u)=N_{ik}(u)-\int_{0}^{u}\widehat{\lambda}_{ik}(u|\overline{\bH}_{iu})Y_{ik}(u)du.
\end{align*}
Note that the times scale defined for $\lambda_{k}(u|\overline{\bH}_{iu})$ is different from the time scale of $\overline{\bH}_{iu}$, so that we may need to re-scale $\overline{\bH}_{iu}$.\\
\textbf{Step 2.} Using the data $(X_{i}, \Delta_{i}, \overline{H}_{i,X_{i}})$ to fit a time-dependent Cox model (regarding $C_{i}$ as time to event and $\tau_{i}$ as time to censoring): $\lambda_{iC}(u|\overline{\bH}_{i,u})=\lambda_{0C}(u)\exp\{\gamma_{C}^{\intercal}g_{C}(u,\overline{\bH}_{i,u})\}$ to obtain the estimate of $\gamma_{C}$: $\widehat{\gamma}_{C}$. Then, we can obtain the estimated survival of censoring:
\begin{align*}
    \widehat{S}_{Ci}(u|\overline{\bH}_{u})=\displaystyle \prod_{V_{i,K_{i}}\leqslant s\leqslant u}\left[1-\widehat{\lambda}_{0C}(s)\exp\{\widehat{\gamma}_{C}^{\intercal}g_{C}(s,\overline{\bH}_{i,s})\}\text{d}s\right].
\end{align*}
Let $N_{iC}(u)=\mathbbm{1}(C_{i}\leqslant u, \Delta_{i}=0)$ and $Y_{iC}(u)=\mathbbm{1}(C_{i}\geqslant u)$ be the counting process and at-risk process of observing censoring respectively, then $\widehat{\gamma}_{C}$ in the above equation can be obtained using the Breslow estimator 
\begin{align*}
    \widehat{\lambda}_{0C}(u)du=\frac{\sum_{i=1}^{n}\text{d}N_{iC}(u)}
    {\sum_{i=1}^{n}\exp\{\widehat{\gamma}_{C}g_{C}(u,\overline{\bH}_{i,u})\}Y_{iC}(u)}.
\end{align*}
\textbf{Step 3.} We solve the following estimating equation to obtain our estimator $\bpsi^{*}$
\begin{align}
\label{final EE}
    \Pbb_{n}\left\{\frac{\Delta_{i}}{\widehat{S}_{Ci}(\tau_{i}|\overline{\bH}_{\tau_{i}})}\sum_{k=1}^{K_{1}} \int c(\overline{\bH}_{i, V_{k-1}+w+u})\left[U_{i}(\bpsi)-\Ebb\left\{U_{i}(\bpsi) |\overline{\bH}_{i, V_{k-1}+w+u}, T_{ik}\geqslant u;\widehat{\xi}\right\}\right] \text{d}\widehat{M}_{ik}(u)\right\}=0,
\end{align}
where we estimate $\Ebb\left\{U_{i}(\bpsi) |\overline{\bH}_{i, V_{k-1}+w+u}, T_{ik}\geqslant u;\xi\right\}$ by regressing $\frac{\Delta_{i}}{\widehat{S}_{Ci}(\tau_{i}|\overline{\bH}_{\tau_{i}})}U_{i}(\bpsi)$ on $(X_{0}, \bL_{u}, u)$ restricted to subjects with $T_{ik}\geqslant u$. The estimating equation (\ref{final EE}) is continuously differentiable on $\bpsi$ and thus can be generally solved using a Newton-Raphson procedure \citep{atkinson2008introduction}.
\section{Discussion}
In this work, taking all the treatment dispensation information into account, we present how to robustly and efficiently estimate the effects of intermittent treatments. The dependent censoring is tackled with the inverse probability of censoring weighting. The proposed estimators are doubly robust and semiparametrically efficient. 

However, we did not take treatment switching into account and the methods cannot estimate the effects of multiple concurrent treatments. In addition, the presence of instrumental variables may result in a biased estimation of the treatment effects in finite samples. We leave these to the future work

\bigskip


\clearpage
\textbf{\large\bf Supplementary material for ``Continuous-time structural failure time model for intermittent treatment''}
\setcounter{page}{1}
\renewcommand{\thepage}{S\arabic{page}}
\begin{appendices}
\numberwithin{equation}{section}
\renewcommand{\theequation}{\thesection.\arabic{equation}}
\numberwithin{table}{section}
\renewcommand{\thetable}{\thesection.\arabic{table}}
\numberwithin{figure}{section}
\renewcommand{\thefigure}{\thesection.\arabic{figure}}
\numberwithin{assumption}{section}
\renewcommand{\theassumption}{\thesection.\arabic{assumption}}
\section{Illustration of the counting process}
    \label{Figure}
\begin{figure}[h]
    \centering
    \begin{tikzpicture}[scale = 1.75]
\draw[black, thick] (0,-0.5) -- (8.5,-0.5);
\draw[black, thick] (0,-0.5) -- (0,2);
\filldraw[black] (0,-0.5) circle (1pt) node[anchor=north] {\scriptsize{$0=V_{i,0}$}};
\filldraw[red] (0.8,2.8) node[anchor=north] {\scriptsize{On}};
\filldraw[blue] (0.8,2.4)  node[anchor=north] {\scriptsize{Off}};
\draw[red, thick] (1.1,2.65) -- (1.5,2.65);
\draw[blue, thick] (1.1,2.25) -- (1.5,2.25);

\draw[red, thick] (0,1) -- (2,1);
\draw[blue, thick] (2,1) -- (4,1);
\draw[red, thick] (4,1) -- (6,1);
\draw[blue, thick] (6,1) -- (7,1);
\draw[red, thick] (7,1) -- (8,1);
\draw [decorate,decoration={brace},xshift=0pt,yshift=0pt]
(0,1.1) -- (2,1.1) node [black,midway,xshift=0.1pt,yshift=5pt] 
{\scriptsize {$w$}};
\draw [decorate,decoration={brace},xshift=0pt,yshift=0pt]
(4,1.1) -- (6,1.1) node [black,midway,xshift=0.1pt,yshift=5pt] 
{\scriptsize {$w$}};

\draw [decorate,decoration={brace},xshift=-1pt,yshift=0pt]
(0.5,2.2) -- (0.5,2.7) node [black,midway,xshift=-1cm] 
{\scriptsize Treatment};
\filldraw[black] (8.5,-0.5) node[anchor=north] {\scriptsize{Time}};
\filldraw[black] (-0.7,1.2) node[anchor=north] {\scriptsize{Subject $i$}};

\draw [decorate,decoration={brace},xshift=0pt,yshift=0pt]
(2,1.1) -- (4,1.1) node [black,midway,xshift=0.1pt,yshift=5pt] 
{\scriptsize {$T_{i,1}$}};
\draw [decorate,decoration={brace},xshift=0pt,yshift=0pt]
(6,1.1) -- (7,1.1) node [black,midway,xshift=0.1pt,yshift=5pt] 
{\scriptsize {$T_{i,2}$}};

\draw[black, dashed] (2,-0.5) -- (2,1);
\draw[blue, thick] (2,0.25) -- (3,0.25);
\draw [decorate,decoration={brace},xshift=0pt,yshift=0pt]
(2,0.3) -- (3,0.3) node [black,midway,xshift=0.1pt,yshift=5pt] 
{\scriptsize {$u$}};
\filldraw[black] (3.1,0.25)  node[anchor=west] {\scriptsize{$N_{i,1}(u)=0$}};

\filldraw[black] (4,-0.5) circle (1pt) node[anchor=north] {\scriptsize{$V_{i,1}$}};
\filldraw[black] (7,-0.5) circle (1pt) node[anchor=north] {\scriptsize{$V_{i,2}$}};
\filldraw[black] (2,-0.5) circle (1pt) node[anchor=north] {\scriptsize{$V_{i,0}+w$}};

\end{tikzpicture}
    \caption{Illustration of the defined gaping time $T_{ik}$ and counting process $N_{ik}(u)$ for the subject $i$.}
\end{figure}
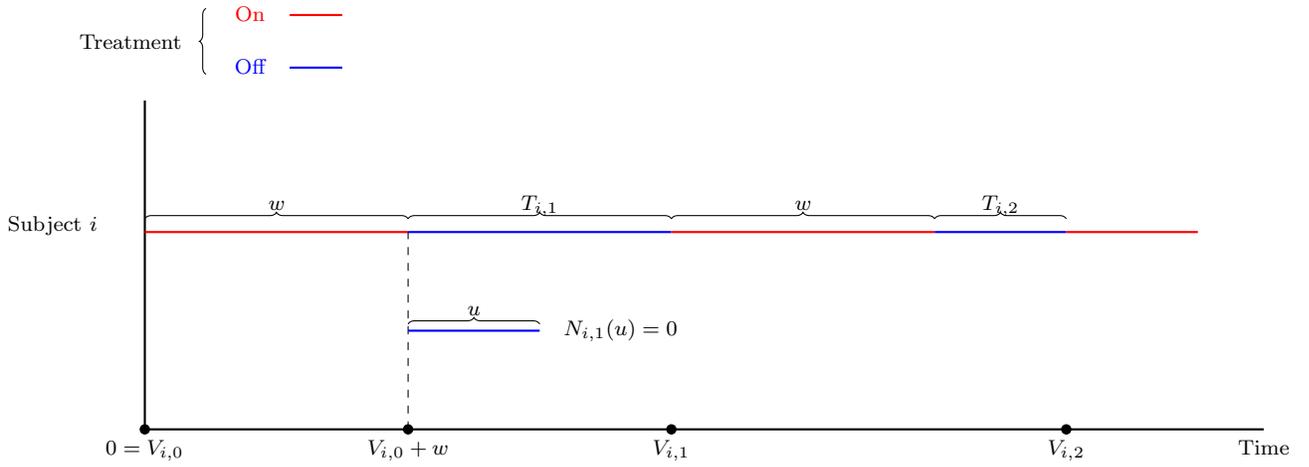

\section{Proof of (\ref{treatment switching hazard equivalence})}
\label{after Assumption 3 appendix}
Because $\lambda_{ik}\left(u | \overline{\bH}_{i, V_{k-1}+w+u}\right)=\lambda_{ik}\left(u | \overline{\bH}_{i, V_{k-1}+w+u}, U_{i}\right)$, so it is sufficient to show that $$\lambda_{ik}\left(u | \overline{\bH}_{i, V_{k-1}+w+u}, U_{i}\right)=\\ \lambda_{ik}\left\{u |\overline{\bH}_{i, V_{k-1}+w+u}, U_{i}(\bpsi^{*})\right\}.$$ Notice that the distribution of $(\overline{\bH}_{i, V_{k-1}+w+u}, U_{i})$ and $\{\overline{\bH}_{i, V_{k-1}+w+u}, U_{i}(\bpsi^{*})\}$ are the same according to Model (\ref{original model}), we have
\begin{align*}
&\lambda_{ik}\left(u | \overline{\bH}_{i, V_{k-1}+w+u}, U_{i}\right)\\
=&\lim _{\text{d}u \rightarrow 0} \text{d}u^{-1} P\left(u \leqslant T_{ik}<u+\text{d}u | T_{ik} \geqslant u, \overline{\bH}_{i, V_{k-1}+w+u}, U_{i}\right) \\
=& \lim _{\text{d}u \rightarrow 0} \text{d}u^{-1} 
\frac{P\left(u \leqslant T_{ik}<u+\text{d}u,  T_{ik} \geqslant u, \overline{\bH}_{i, V_{k-1}+w+u}, U_{i}\right)} {P\left( T_{ik} \geqslant u, \overline{\bH}_{i, V_{k-1}+w+u}, U_{i}\right)}\\
=& \lim _{\text{d}u \rightarrow 0} \text{d}u^{-1} 
\frac{P\left(u \leqslant T_{ik}<u+\text{d}u| T_{ik} \geqslant u, \overline{\bH}_{i, V_{k-1}+w+u}\right)} {P\left(U_{i}| T_{ik} \geqslant u, \overline{\bH}_{i, V_{k-1}+w+u}\right)}\\
&\frac{P\left(U_{i}|u \leqslant T_{ik}<u+\text{d}u,\overline{\bH}_{i, V_{k-1}+w+u}\right)P\left(T_{ik} \geqslant u, \overline{\bH}_{i, V_{k-1}+w+u}\right)}{P\left( T_{ik} \geqslant t, \overline{\bH}_{i, V_{k-1}+w+u}\right)}\\
=& \lim _{\text{d}u \rightarrow 0} \text{d}u^{-1} \frac{P\left(U_{i} | u \leqslant T_{ik}<u+\text{d}u, \overline{\bH}_{i, V_{k-1}+w+u}\right) P\left(u \leqslant T_{ik}<u+\text{d}u | T_{ik} \geqslant u, \overline{\bH}_{i, V_{k-1}+w+u}\right)}{P\left(U _{i}| T_{ik} \geqslant u, \overline{\bH}_{i, V_{k-1}+w+u}\right)} \\
=& \lim _{\text{d}u \rightarrow 0} \text{d}u^{-1} \frac{P\left\{U_{i}\left(\bpsi^{*}\right) | u \leqslant T_{ik}<u+\text{d}u, \overline{\bH}_{i, V_{k-1}+w+u}\right\} P\left(u \leqslant T_{ik}<u+\text{d}u | T_{ik} \geqslant u, \overline{\bH}_{i, V_{k-1}+w+u}\right)}{P\left\{U\left(\bpsi^{*}\right) | T_{ik} \geqslant u,  \overline{\bH}_{i, V_{k-1}+w+u}\right\}} \\
=& \lim _{\text{d}u \rightarrow 0} \text{d}u^{-1} P\left\{u \leqslant T_{ik}<u+\text{d}u | T_{ik} \geqslant u, \overline{\bH}_{i, V_{k-1}+w+u}, U_{i}\left(\bpsi^{*}\right)\right\} \\
=& \lambda_{ik}\left\{u | \overline{\bH}_{i, V_{k-1}+w+u}, U_{i}\left(\bpsi^{*}\right)\right\}
\end{align*}

\section{Proof of (\ref{martingale})}
\label{martingale proof}
The details of the proof is shown in Theorem 1.3.1 and 1.3.2 in \citep{fleming1991counting}. Here, we provide a heuristic proof.\\
To show (\ref{martingale}) is indeed a martingale, it is sufficient to show that $\Ebb\{dM_{ik}(t)|\mathscr{F}_{i,V_{k-1}+w+u}\}=0$.
\begin{align*}
    &\Ebb\{\text{d}M_{ik}(u)|\mathscr{F}_{i,V_{k-1}+w+u}\}\\
    =&\Ebb\{\text{d}N_{ik}(u)-\lambda_{ik}(u|\overline{\bH}_{i,V_{k-1}+w+u})Y_{ik}(u)\text{d}u|\mathscr{F}_{i,V_{k-1}+w+u}\}\\
    =&\Ebb\{\mathbbm{1}(u=T_{ik} )\text{d}u|\mathscr{F}_{i,V_{k-1}+w+u}\}\\
    -&\Ebb\{\lim _{\text{d}u \rightarrow 0} \text{d}u^{-1} P\left(u \leqslant T_{ik}<u+\text{d}u | T_{ik} \geqslant u, \overline{\bH}_{i, V_{k-1}+w+u}, U_{i}(\bpsi^{*})\right)\mathbbm{1}(u\leqslant T_{ik})\text{d}u|\mathscr{F}_{i,V_{k-1}+w+u}\}\\
    =&\Ebb\{\mathbbm{1}(u=T_{ik} )\text{d}u|\mathscr{F}_{i,V_{k-1}+w+u}\}\\
    -&\Ebb\{\lim _{\text{d}u \rightarrow 0} \text{d}u^{-1} P\left(u \leqslant T_{ik}<u+\text{d}u | \overline{\bH}_{i, V_{k-1}+w+u}, U_{i}(\bpsi^{*})\right)\text{d}u|\mathscr{F}_{i,V_{k-1}+w+u}\}\\
     =&\Ebb\{\mathbbm{1}(u=T_{ik} )\text{d}u|\mathscr{F}_{i,V_{k-1}+w+u}\}-\Ebb\{\mathbbm{1}(u=T_{ik} )\text{d}u|\mathscr{F}_{i,V_{k-1}+w+u}\}\\
     =&0
\end{align*}

\section{A Lemma}
\label{martingale_lemma}
We provide a lemma for the martingale process, which is used in the later proof.
\begin{lemma1}
\label{Lemma1}
Consider the martingales $M_{ik}(u), \forall k$ defined in (\ref{martingale}), according to Theorem 2.6.2, Lemma 2.6.1 of \citep{fleming1991counting} and II.4.1 in \citep{andersen1993statistical}, if arbitrary functions $f_{k}(\cdot)$ and $g_{k}(\cdot)$ are bounded $\mathscr{F}_{i, V_{i,k_1}+w+u}$-predictable processes on $[0, \infty)$ with Assumption \ref{independent gapping time} holds, then both $\int f_{k}(\cdot)\textnormal{d}M_{ik}(u)$ and $M_{f}=\sum_{k=1}^{K_{i}}\int f_{k}(\cdot)\textnormal{d}M_{ik}(u)$ are martingales. Specifically, $\Ebb(M_{f})=0$ and $\forall u\in[0, \infty)$, the covariance of $M_{f}$ and $M_{g}$ is 
\begin{align*}
    Cov(M_{f}, M_{g})
    =&\Ebb\left\{\sum_{k=1}^{K_{i}}\int f_{k}(\cdot)\textnormal{d}M_{ik}(u)\times\sum_{k=1}^{K_{i}}\int g_{k}(\cdot)\textnormal{d}M_{ik}(u)\right\}\\
    =&\Ebb\left\{\sum_{k=1}^{K_{i}}\sum_{l=1}^{K_{i}}<\int f_{k}(\cdot)\textnormal{d}M_{ik}(u),\int g_{l}(\cdot)dM_{il}(u)>\right\}\\
    =& \Ebb\left\{\sum_{k=1}^{K_{i}}\sum_{l=1}^{K_{i}}\int f_{k}(\cdot)g_{l}(\cdot)d<M_{ik}(u),M_{il}(u)>\right\}\\
    =& \Ebb\left\{\sum_{k=1}^{K_{i}}\int f_{k}(\cdot)g_{k}(\cdot)d<M_{ik}(u),M_{ik}(u)>\right\}\\
    =&\sum_{k=1}^{K_{i}}\Ebb\Bigg[\int f_{k}(\cdot)g_{k}(\cdot)Y_{ik}(u)\lambda_{ik}\left(u | \overline{\bH}_{i, V_{k-1}+w+u}\right)\\
    &\left\{1-\Delta\int \lambda_{ik}\left(u | \overline{\bH}_{i, V_{k-1}+w+u}\right) \text{d}u\right\}\textnormal{d}u\Bigg]\\
    =&\sum_{k=1}^{K_{i}}\int\Ebb\left\{ f_{k}(\cdot)g_{k}(\cdot)Y_{ik}(u)\right\}\lambda_{ik}\left(u | \overline{\bH}_{i, V_{k-1}+w+u}\right\}\\
    &\left\{1-\Delta\int \lambda_{ik}\left(u | \overline{\bH}_{i, V_{k-1}+w+u}\right) \text{d}u\right\}\textnormal{d}u
\end{align*}
where $<M_1, M_2>$ is the compensator of $M_{1}M_{2}$.
\end{lemma1}

\section{Proof of Theorem \ref{theorem 1}}
\label{theorem 1 appendix}
We first look into the likelihood function based on a single observable $\bO_{i}$. To decompose the likelihood, we define $0=v_{(0)}<v_{(1)}<\dots<v_{(M)}$ as an ordered and fined increasing time sequence on the time line during the follow-up, such that the values of the covariates and treatment remain the same between two successive time points, and $v_{(M)}$ is the time right before $\tau_{i}$. Therefore, we can write $\overline{\bH}_{iv_{(m)}}$ as $\{\overline{\bL}_{iv_{(m)}},\overline{A}_{iv_{(m-1)}}\}$.\\ 
Since the map from $\bO_{i}$ to $\left\{U_{i}(\bpsi^{*}),\overline{\bH}_{i, V_{k-1} +w+u}\right\}$ is one-to-one with  Jacobian determinant 
\begin{align*}
\frac{\partial U_{i}(\bpsi^{*})}{\partial \tau_{i}}=  
     \exp \left[\left\{\psi_{1}  +\bpsi_{2}^{j* \intercal} g\left(\bL_{iu}\right)\right\}A_{iu}\right],
\end{align*}
we have
\begin{align*}
&f_{\bO_{i}}(\bO_{i};\bpsi^{*} )\\
=&\left\{\frac{\partial U_{i}(\bpsi^{*})}{\partial \tau_{i}}\right\}
f_{\left\{U_{i}(\bpsi^{*}),\overline{\bH}_{i, \tau_{i}}\right\}}\left\{U_{i}(\bpsi^{*}),\overline{\bH}_{i, \tau_{i}} \right\}\\
=&\left\{\frac{\partial U_{i}(\bpsi^{*})}{\partial \tau_{i}}\right\}
f\left\{U_{i}(\bpsi^{*}) \right\}f\left\{\overline{\bH}_{i, \tau_{i}}|U_{i}(\bpsi^{*})  \right\}\\
=&\left\{\frac{\partial U_{i}(\bpsi^{*})}{\partial \tau_{i}}\right\}
f_{1}\left\{U_{i}(\bpsi^{*}) \right\}\prod_{m=1}^{M}f_2\left\{\bL_{iv_{(m)}}|\overline{\bL}_{iv_{(m-1)}},\overline{A}_{iv_{(m-1)}},U_{i}(\bpsi^{*}) \right\}\\
& \times\prod_{m=1}^{M}f_3\left\{A_{iv_{(m)}}|\overline{\bL}_{iv_{(m)}},\overline{A}_{iv_{(m-1)}},U_{i}(\bpsi^{*}) \right\}\\
=&\left\{\frac{\partial U_{i}(\bpsi^{*})}{\partial \tau_{i}}\right\}
f_{1}\left\{U_{i}(\bpsi^{*}) \right\}\prod_{m=1}^{M}f_2\left\{\bL_{iv_{(m)}}|\overline{\bH}_{iv_{(m)}}, A_{iv_{(m-1)}},U_{i}(\bpsi^{*}) \right\}\\
& \times\prod_{m=1}^{M}f_3\left\{A_{iv_{(m)}}|\overline{\bH}_{iv_{(m)}},U_{i}(\bpsi^{*}) \right\}\\
=&\left\{\frac{\partial U_{i}(\bpsi^{*})}{\partial \tau_{i}}\right\}
f_{1}\left\{U_{i}(\bpsi^{*}) \right\}\prod_{m=1}^{M}f_2\left\{\bL_{iv_{(m)}}|\overline{\bH}_{iv_{(m)}}, A_{iv_{(m-1)}},U_{i}(\bpsi^{*}) \right\}\\
& \times\prod_{m=1}^{M}f_3\left\{A_{iv_{(m)}}|\overline{\bH}_{iv_{(m)}} \right\} \hspace{0.1cm} \textnormal{by Assumption \ref{NUC} and (\ref{treatment switching hazard equivalence})}\\
=&\left\{\frac{\partial U_{i}(\bpsi^{*})}{\partial \tau_{i}}\right\}
f_{1}\left\{U_{i}(\bpsi^{*}) \right\}\prod_{m=1}^{M}f_2\left\{\bL_{iv_{(m)}}|\overline{\bH}_{iv_{(m)}}, A_{iv_{(m-1)}},U_{i}(\bpsi^{*}) , \tau_{i}>v_{(m)}\right\}\\
& \times\prod_{m=1}^{M}f_3\left\{A_{iv_{(m)}}|\overline{\bH}_{iv_{(m)}}, \tau_{i}>v_{(m)} \right\}
\end{align*}
where $f$ denotes a valid density or likelihood function which satisfies $\int f \mathrm{d} \nu$=1, where $\nu$ is the dominating measure of $f$. The last equation follows because $f_2$ and $f_3$ only valid when $\tau_{i}>v_{(m)}$ due to the nature of survival data: the values of $\bL_{iv_{(m)}}$ and $A_{iv_{(m)}}$ are not available when $\tau_{i}\leqslant v_{(m)}$.\\
The above model is said to be a semiparametric model in the sense that $f_1$, $f_2$ and $f_3$ are not only specified by finite parameters of interest $\bpsi^{*}$, but also characterized by infinite-dimensional nuisance parameters, which we denote as $\btheta=(\btheta_1, \btheta_2, \btheta_3)$. A model is called the parametric submodels $f(\cdot;\btheta)$ of the model $f(\cdot)$, if it contains the true model of $f(\cdot)$ when it is evaluated at $\btheta=0$. We define the models 
\begin{align*}
    &f_{1}\left\{U_{i}(\bpsi^{*});\btheta_{1} \right\},\\ &f_2\left\{\bL_{iv_{(m)}}|\overline{\bH}_{iv_{(m)}}, A_{iv_{(m-1)}}, U_{i}(\bpsi^{*}),\tau_{i}>v_{(m)};\btheta_{2}\right\},\\ &f_3\left\{A_{iv_{(m)}}|\overline{\bH}_{iv_{(m)}}, \tau_{i}>v_{(m)} ;\btheta_{3}\right\}
\end{align*}
as the parametric submodels of $f_{1}\left\{U_{i}(\bpsi^{*}) \right\}$, $f_2\left\{\bL_{iv_{(m)}}|\overline{\bH}_{iv_{(m)}},\\ A_{iv_{(m-1)}},U_{i}(\bpsi^{*}) , \tau_{i}>v_{(m)}\right\}$, and $f_3\left\{A_{iv_{(m)}}|\overline{\bH}_{iv_{(m)}}, \tau_{i}>v_{(m)} \right\}$ respectively.\\
To find the orthogonal complement of nuisance tangent space, let us first characterize the nuisance tangent space $\Lambda$ with respect to the $\btheta$. Assume $\btheta_i, i=1, 2, 3$ is variationally independent to each other, then $\Lambda$ is defined as mean-square closure of $\Lambda_i, i=1, 2, 3$, where $\Lambda_i$ is the parametric submodel nuisance tangent space of $\btheta_i, i=1, 2, 3$, and $\Lambda=\Lambda_1\oplus\Lambda_2\oplus\Lambda_3$, where $\oplus$ is the direct sum \citep{tsiatis2007semiparametric}.\\
Since $f_1$ and $f_2$ are a marginal and a conditional nonparametric model, respectively without any other restriction, following Section 4.4 of \citep{tsiatis2007semiparametric}, we know that the $\Lambda_1$ and $\Lambda_2$ are a set of all mean-zero vectors $\alpha\{U_{i}(\bpsi^{*})\}$, and a set of score vectors $S\{\bL_{iv_{(m)}}, \overline{\bH}_{iv_{(m)}}, A_{iv_{(m-1)}},U_{i}(\bpsi^{*})\}=S\{\overline{\bH}_{iv_{(m)}}, U_{i}(\bpsi^{*})\}$ respectively. We thus express the parametric submodel nuisance tangent spaces with respect to $\btheta_1$ and $\btheta_2$ as
\begin{align*}
    \Lambda_1=&\left\{\alpha\{U_{i}(\bpsi^{*})\}\in\Rbb^{ p}: \Ebb\left[\alpha\{U_{i}(\bpsi^{*})\right]=0\right\},\\
    \Lambda_2=&\left\{S\{\overline{\bH}_{iv_{(m)}}, U_{i}(\bpsi^{*})\}\in \Rbb^{ p}:\Ebb\left[S\{\overline{\bH}_{iv_{(m)}}, U_{i}(\bpsi^{*})\}|\overline{\bH}_{iv_{(m)}}, A_{iv_{(m-1)}},U_{i}(\bpsi^{*})\right]=0\right\}
\end{align*}
To derive $\Lambda_3$, we first notice that because the treatment dispensation process can be fully specified by the event of multiple times of prescription pickup. This is because $A_{iu}$ is 1 in $(V_{i,k-1}, V_{i,k-1}+w]$ and 0 otherwise. Therefore, we can express $\prod_{m=1}^{M} f_3  \left\{A  _{iv_{(m)}}|\overline{\bH}_{iv_{(m)}}, \tau_{i}>v_{(m)} ;\btheta_{3}  \right\}$ equivalently as the likelihood based on the hazard of refilling the prescription: $\lambda_{ik}\left(u | \overline{\bH}_{i, V_{k-1} +w+u},\btheta_{3} \right)$. Since we are only interested in the nuisance parameter $\btheta_{3}$ that describes the treatment model, rather than the model of time event, a likelihood that contributed by $\lambda_{ik}\left(u | \overline{\bH}_{i, V_{k-1} +w+u},\btheta_{3} \right)$ is sufficient to be considered to derive the nuisance tangent space with respect to $\btheta_{3}$.  Due to Assumption \ref{NUC} (similar to recurrent event based on gaping time),
\begin{align}
\label{likelihood}
\begin{split}
& \prod_{m=1}^{M} f_3  \left\{A  _{iv_{(m)}}|\overline{\bH}_{iv_{(m)}}, \tau_{i}>v_{(m)} ;\btheta_{3}  \right\}\\
\propto&\prod_{k=1}^{K_{i}}\lambda_{ik}\left(t_{ik} | \overline{\bH}_{i, V_{k-1} +w+u};\btheta_{3}\right) \exp\left\{-\int_{0}^{t_{ik}}\lambda_{ik}\left(u| \overline{\bH}_{i, V_{k-1} +w+u};\btheta_{3}\right\}\text{d}u\right\}
\end{split}
\end{align}
Assume a parametric submodel 
\begin{align}
\label{submodel}
\lambda_{ik}\left(t_{ik} | \overline{\bH}_{i, V_{k-1} +w+u};\btheta_{3}\right)=\lambda_{0ik}\left(t_{ik} | \overline{\bH}_{i, V_{k-1} +w+u}\right)\exp\{\btheta_{3}^{\intercal} h(t_{ik},\overline{\bH}_{i, V_{k-1} +w+u})\},
\end{align}
where $\lambda_{0ik}\left(t_{ik} | \overline{\bH}_{i, V_{k-1} +w+u}\right)$ denotes the true model of the hazard and $h(t_{ik},\overline{\bH}_{i, V_{k-1} +w+u})$ an arbitrary $ p$-dimensional function. Plug (\ref{submodel}) into log of (\ref{likelihood}), take derivative with respect to $\btheta_3$ and evaluate it at $\btheta_3=0$, we obtained the score
\begin{align*}
    S_3=&\sum_{k=1}^{K_{i}}\left[h(t_{ik},\overline{\bH}_{i, V_{k-1} +w+u})-\int_{0}^{t_{ik}}\lambda_{0ik}(u | \overline{\bH}_{i, V_{k-1} +w+u})h(u,\overline{\bH}_{i, V_{k-1} +w+u})\text{d}u\right]\\
    =&\sum_{k=1}^{K_{i}}\left[\int h(u,\overline{\bH}_{i, V_{k-1} +w+u})\text{d}N_{ik}(u)-\int \lambda_{0ik}(u | \overline{\bH}_{i, V_{k-1} +w+u})h(u,\overline{\bH}_{i, V_{k-1} +w+u})Y_{ik}(u)\text{d}u\right]\\
    =&\sum_{k=1}^{K_{i}}\int h(u,\overline{\bH}_{i, V_{k-1} +w+u}) \text{d}M_{ik}(u),
\end{align*}
which is the sum of stochastic integral of $M_{ik}(u)$ over $h(u,\overline{\bH}_{i, V_{k-1} +w+u})$. Thus the parametric submodel nuisance tangent with respect to $\btheta_{3}$ is 
\begin{align*}
    \Lambda_{3}=\left\{\sum_{k=1}^{K_{i}}\int h(u,\overline{\bH}_{i, V_{k-1} +w+u}) \text{d}M_{ik}(u):h(u,\overline{\bH}_{i, V_{k-1} +w+u})\in\Rbb^{ p}  \right\}.
\end{align*}
It is not trivial to directly derive the orthogonal space of $\Lambda$ by finding the orthogonal space of three parametric submodel nuisance tangent spaces, instead, we can consider a nonparametric model first. The only constrain of the model is $\lambda_{ik}(u | \bO_{i}, U_{i})
=\lambda_{ik}\left(u | \overline{\bH}_{i,V_{k-1} +w+u}\right)$, which results in a different $\Lambda_{3}$. If we relax that restriction, the nonparametric model nuisance tangent space with respect to $\btheta_{3}$ would be 
\begin{align*}
    \Lambda_{3}^{*}=\left[\sum_{k=1}^{K_{i}}\int h\left\{u, U_{i}(\bpsi^{*}), \overline{\bH}_{i, V_{k-1} +w+u}\right\} \text{d}M_{ik}(u):\left\{u,U_{i}(\bpsi^{*}), \overline{\bH}_{i, V_{k-1} +w+u}\right\}\in\Rbb^{ p}  \right].
\end{align*}
It is known that the nuisance tangent space for a nonparametric model is the entire Hilbert space which contains $\Lambda$, so $\mathcal{H}=\Lambda_{1}\oplus\Lambda_{2}\oplus\Lambda_{3}^{*}$. That implies that $\Lambda_{3}\subset\Lambda_{3}^{*}$. Since the orthogonal complement $\Lambda^{\perp}$ is the complement space that is orthogonal to $\Lambda_{1}\oplus\Lambda_{2}$, so that $\Lambda_{3}\subset\Lambda_{3}^{*}$. As a result, $\Lambda^{\perp}$ consists of elements of $\Lambda_{3}^{*}$ that is orthogonal to $\Lambda_{3}$. Therefore, the elements of  $\Lambda^{\perp}$ should be the resi\text{d}ual after projecting an arbitrary element of $\Lambda_{3}^{*}$ onto $\Lambda_{3}$. In order to characterize the resi\text{d}ual, we have to derive $h^{*}(u, \overline{\bH}_{i, V_{k-1} +w+u})$ such that the resi\text{d}ual
\begin{align*}
    \left[\sum_{k=1}^{K_{i}}\int h\left\{u,U_{i}(\bpsi^{*}), \overline{\bH}_{i, V_{k-1} +w+u}\right\}\text{d}M_{ik}(u)-\sum_{k=1}^{K_{i}}\int h^{*}(u, \overline{\bH}_{i, V_{k-1} +w+u})\text{d}M_{ik}(u)\right]
\end{align*}
is orthogonal to all the elements of $\Lambda_{3}$. That is, $\forall h(u,\overline{\bH}_{i, V_{k-1} +w+u})$, 
\begin{align}
\label{EE1}
\begin{split}
    \Ebb\Bigg(\sum_{k=1}^{K_{i}}&\int\Big[h\left\{u,U_{i}(\bpsi^{*}), \overline{\bH}_{i, V_{k-1} +w+u}\right\} -h^{*}(u, \overline{\bH}_{i, V_{k-1} +w+u})\Big]\text{d}M_{ik}(u)\\
    &\times\sum_{k=1}^{K_{i}}\int h\left\{u, \overline{\bH}_{i, V_{k-1} +w+u}\right\} \text{d}M_{ik}(u)\Bigg)=0.
\end{split}
\end{align}
By Lemma \ref{Lemma1}, (\ref{EE1}) becomes
{\small
\begin{align*}
\begin{split}
   & \sum_{k=1}^{K_{i}}\int\Ebb\Big(\Big[h\left\{u,U_{i}(\bpsi^{*}), \overline{\bH}_{i, V_{k-1} +w+u}\right\} -h^{*}(u, \overline{\bH}_{i, V_{k-1} +w+u})\Big]h\left\{u, \overline{\bH}_{i, V_{k-1} +w+u}\right\}
   Y_{ik}(u)\Big)\\
   &\lambda_{ik}\left(u | \overline{\bH}_{i, V_{k-1}+w+u}\right)\left\{1-\Delta\int \lambda_{ik}\left(u | \overline{\bH}_{i, V_{k-1}+w+u}\right) \text{d}u\right\}\text{d}u\\
   =& \sum_{k=1}^{K_{i}}\int\Ebb\Bigg\{\Ebb\Big(\Big[h\left\{u,U_{i}(\bpsi^{*}), \overline{\bH}_{i, V_{k-1} +w+u}\right\} -h^{*}(u, \overline{\bH}_{i, V_{k-1} +w+u})\Big]h\left\{u, \overline{\bH}_{i, V_{k-1} +w+u}\right\}
   Y_{ik}(u)|\overline{\bH}_{i, V_{k-1} +w+u}\Big)\Bigg\}\\
   &\lambda_{ik}\left(u | \overline{\bH}_{i, V_{k-1}+w+u}\right)\left\{1-\Delta\int \lambda_{ik}\left(u | \overline{\bH}_{i, V_{k-1}+w+u}\right) \text{d}u\right\}\text{d}u\\
   =&\sum_{k=1}^{K_{i}}\int\Ebb\Bigg\{\Ebb\Big(\Big[h\left\{u,U_{i}(\bpsi^{*}), \overline{\bH}_{i, V_{k-1} +w+u}\right\} -h^{*}(u, \overline{\bH}_{i, V_{k-1} +w+u})\Big]Y_{ik}(u)|\overline{\bH}_{i, V_{k-1} +w+u}\Big)
   h\left\{u, \overline{\bH}_{i, V_{k-1} +w+u}\right\}\Bigg\}\\
   &\lambda_{ik}\left(u | \overline{\bH}_{i, V_{k-1}+w+u}\right)\left\{1-\Delta\int \lambda_{ik}\left(u | \overline{\bH}_{i, V_{k-1}+w+u}\right) \text{d}u\right\}\text{d}u\\
   =&0
\end{split}
\end{align*}}
for any $h\left\{u, \overline{\bH}_{i, V_{k-1} +w+u}\right\}$. Since $h\left\{u, \overline{\bH}_{i, V_{k-1} +w+u}\right\}$ is arbitrary, $\lambda_{ik}\left(u | \overline{\bH}_{i, V_{k-1}+w+u}\right)$ is positive, and $1-\Delta\int \lambda_{ik}\left(u | \overline{\bH}_{i, V_{k-1}+w+u}\right) \text{d}u\in(0,1)$, we must have $\forall k$,
\begin{align}
\label{first estimating equation}
    \Ebb\Big(\Big[h\left\{u,U_{i}(\bpsi^{*}), \overline{\bH}_{i, V_{k-1} +w+u}\right\} -h^{*}(u, \overline{\bH}_{i, V_{k-1} +w+u})\Big]Y_{ik}(u)|\overline{\bH}_{i, V_{k-1} +w+u}\Big)=0.
\end{align}
Solving (\ref{first estimating equation}), we obtain
\begin{align*}
    \Ebb\Big[h\left\{u,U_{i}(\bpsi^{*}), \overline{\bH}_{i, V_{k-1} +w+u}\right\} Y_{ik}(u)|\overline{\bH}_{i, V_{k-1} +w+u}\Big]=
    h^{*}(u, \overline{\bH}_{i, V_{k-1} +w+u})\Ebb\{Y_{ik}(u)|\overline{\bH}_{i, V_{k-1} +w+u}\},
\end{align*}
or 
\begin{align*}
    h^{*}(u, \overline{\bH}_{i, V_{k-1} +w+u})&=\frac{\Ebb\Big[h\left\{u,U_{i}(\bpsi^{*}), \overline{\bH}_{i, V_{k-1} +w+u}\right\} Y_{ik}(u)|\overline{\bH}_{i, V_{k-1} +w+u}\Big]}{\Ebb\{Y_{ik}(u)|\overline{\bH}_{i, V_{k-1} +w+u}\}}\\
    &=\Ebb\Big[h\left\{u,U_{i}(\bpsi^{*}), \overline{\bH}_{i, V_{k-1} +w+u}\right\} |\overline{\bH}_{i, V_{k-1} +w+u}, T_{ik}\geqslant u\Big].
\end{align*}
Therefore, the space orthogonal to the nuisance tangent space is given by
\begin{align*}
\begin{split}
    \Lambda^{\perp}=\Bigg\{&\sum_{k=1}^{K_{i}}\int\left( h\left\{u,U_{i}(\bpsi^{*}), \overline{\bH}_{i, V_{k-1}+w+u}\right\}-\Ebb\Big[h\left\{u,U_{i}(\bpsi^{*}), \overline{\bH}_{i, V_{k-1}+w+u}\right\} |\overline{\bH}_{i, V_{k-1}+w+u}, T_{ik}\geqslant u\Big]\right)\\
    &\text{d}M_{ik}(u) : h\left\{u,U_{i}(\bpsi^{*}), \overline{\bH}_{i, V_{k-1}+w+u}\right\}\in\Rbb^{ p}\Bigg\}.
\end{split}
\end{align*}

\section{Identification of $\bpsi\in\Rbb^{p}$ under Assumptions 1-4}
\label{identifiability1}
Under Assumption 3, $M_{ik}(u)$ is a $\mathscr{F}_{i,V_{k-1}+w+u}$-predictable zero-mean martingale. Then for all for all bounded and $\mathscr{F}_{i,V_{k-1}+w+u}$-predictable $$c(\overline{\bH}_{i, V_{k-1}+w+u})\left[U_{i}(\bpsi^{*})-\Ebb\left\{U_{i}(\bpsi^{*}) |\overline{\bH}_{i, V_{k-1}+w+u}, T_{ik}\geqslant u\right\}\right]:$$ \begin{align*}
    &\Ebb\left\{\sum_{k=1}^{K_{1}} \int c(\overline{\bH}_{i, V_{k-1}+w+u})\left[U_{i}(\bpsi^{*})-\Ebb\left\{U_{i}(\bpsi^{*}) |\overline{\bH}_{i, V_{k-1}+w+u}, T_{ik}\geqslant u\right\}\right] \text{d}M_{ik}(u)\right\}\\
    =&\Ebb\left\{\sum_{k=1}^{K_{1}} \int c(\overline{\bH}_{i, V_{k-1}+w+u})B(\bpsi^{*})\text{d}M_{ik}(u)\right\}=0.
\end{align*}
Suppose $\bpsi^{*1}$ and $\bpsi^{*2}$ satisfy the above equation, then 
$$\Ebb\left\{\sum_{k=1}^{K_{1}} \int c(\overline{\bH}_{i, V_{k-1}+w+u})\{B(\bpsi^{*1})-B(\bpsi^{*2})\}\text{d}M_{ik}(u)\right\}=0.$$ To reflect the dependent of $B(\bpsi^{1*})-B(\bpsi^{*2})$ on $(\overline{A}_{i,V_{k-1}+w+u},\overline{L}_{i,V_{k-1}+w+u})$, denote 
\begin{align*}
&\phi(\overline{A}_{i,V_{k-1}+w+u},\overline{L}_{i,V_{k-1}+w+u})\\
=&B(\bpsi^{*1})-B(\bpsi^{*2})\\
=&U_{i}(\bpsi^{*1})-\Ebb\left\{U_{i}(\bpsi^{*1}) |\overline{\bH}_{i, V_{k-1}+w+u}, T_{ik}\geqslant u\right\}-
U_{i}(\bpsi^{*2})-\Ebb\left\{U_{i}(\bpsi^{*2}) |\overline{\bH}_{i, V_{k-1}+w+u}, T_{ik}\geqslant u\right\}\\
=&\int_{0}^{\tau_{i}} \exp \left[\left\{\psi_{1}^{*1}+\bpsi_{2}^{*1 \intercal} g\left(\bL_{iu}\right)\right\} A_{iu}\right] \mathrm{d} u-\int_{0}^{\tau_{i}} \exp \left[\left\{\psi_{1}^{*2}+\bpsi_{2}^{*2 \intercal} g\left(\bL_{iu}\right)\right\} A_{iu}\right] \mathrm{d} u+f(\overline{\bH}_{i, V_{k-1}+w+u}),
\end{align*}
where $f(\overline{\bH}_{i, V_{k-1}+w+u})$ is a function of $\overline{\bH}_{i, V_{k-1}+w+u}$. Then we have 
\begin{align*}
\Ebb\left\{\sum_{k=1}^{K_{1}} \int c(\overline{\bH}_{i, V_{k-1}+w+u})\phi(\overline{A}_{i,V_{k-1}+w+u},\overline{L}_{i,V_{k-1}+w+u})\text{d}M_{ik}(u)\right\}=0,
\end{align*} 
which implies that $\forall \overline{H}_{it}, t>0,  \phi(\overline{A}_{i,V_{k-1}+w+u},\overline{L}_{i,V_{k-1}+w+u})$ is independent of $M_{ik}(u)$ conditional on $\overline{H}_{i,V_{k-1}+w+u}, T_{i}\geqslant u$. Therefore, $\phi(\overline{A}_{i,V_{k-1}+w+u},\overline{L}_{i,V_{k-1}+w+u})$ must be independent of $\overline{A}_{i,V_{k-1}+w+u}$, which implies $\bpsi^{*1}$ must equals to $\bpsi^{*2}$, so that $\bpsi^{*}$ is identifiable.

\section{The optimal choice of $c(\overline{\bH}_{i, V_{k-1}+w+u})$}
\label{c_opt}
The estimator obtained by solving the estimating equation( \ref{replaced estimating equation}) may not be the most efficient one. Instead, we investigate the estimating equation by setting the empirical mean of the efficient score function. The efficient score is obtained by projecting a score vector $S_{\bpsi^{*}}(\bO_{i})$ onto the orthogonal complement of nuisance tangent space $\Lambda^{\perp}$. Since $h\left\{u,U_{i}(\bpsi^{*}), \overline{\bH}_{i, V_{k-1}+w+u}\right\}$ is in general intractable, we focus on the reduced class $$h\left\{u,U_{i}(\bpsi^{*}), \overline{\bH}_{i, V_{k-1}+w+u}\right\}=c(\overline{\bH}_{i, V_{k-1}+w+u})\\U_{i}(\bpsi^{*}),$$ within this class, the orthogonal complement of nuisance tangent $\Lambda_{0}^{\perp}$ is
\begin{align*}
    \Lambda_{0}^{\perp}=\Bigg\{G(\bpsi^{*};\bO_{i},c)=
    &\sum_{k=1}^{K_{i}}\int_{0}^{\infty}c(\overline{\bH}_{i, V_{k-1}+w+u})\left[U_{i}(\bpsi^{*})-\Ebb\left\{U_{i}(\bpsi^{*} ) |\overline{\bH}_{i, V_{k-1}+w+u}, T_{ik}\geqslant u\right\}\right] \\ & \text{d}M_{ik}(u):c(\overline{\bH}_{i, V_{k-1}+w+u})\in\Rbb^{p}.    \Bigg\}
\end{align*}
The following theorem gives the resulting efficient score function in this class.
\begin{theorem1}
\label{efficient score}
The efficient score of $\widehat{\bpsi}$ is
\begin{align*}
    S^{\textnormal{eff}}_{\bpsi^{*}}(\bO_{i})
    =&\prod(S_{\bpsi^{*}}(\bO_{i})|\Lambda_{0}^{\perp})\\
    =&-\sum_{k=1}^{K_{i}}\int\left\{1-\Delta\int \lambda_{ik}\left(u | \overline{\bH}_{i, V_{k-1}+w+u}\right) \text{d}u\right\}^{-1}\\
    &\Ebb\Bigg(\frac{\partial}{\partial \bpsi}\left[ U_{i}(\bpsi^{*})-\Ebb\{U_{i}(\bpsi^{*})|\overline{\bH}_{i, V_{k-1}+w+u}, T_{ik}\geqslant u\}\right]|\overline{\bH}_{i, V_{k-1}+w+u}, T_{ik}= u\Bigg)\times\\
    &\left[\textnormal{Var}\{U_{i}(\bpsi^{*})|\overline{\bH}_{i, V_{k-1}+w+u}, T_{ik}\geqslant u\}\right]^{-1}\\
    &\left[U_{i}(\bpsi^{*})-\Ebb\left\{U_{i}(\bpsi^{*}) |\overline{\bH}_{i, V_{k-1}+w+u}, V_{i,k}\geqslant V_{i,k-1}+w+u\right\}\right]\textnormal{d}M_{ik}(u).
\end{align*}
\end{theorem1}
Compared $S^{\textnormal{eff}}_{\bpsi^{*}}(\bO_{i})$ with (\ref{estimating equation}), we thus obtain $c^{\text{opt}}(\overline{\bH}_{i, V_{k-1}+w+u})$ as showed in (\ref{copt}). Next we provide the proof of Theorem \ref{efficient score}.\\
\textit{proof.} We first prove the projection of $S_{\bpsi}(\bO_{i})$ onto $\Lambda_{0}^{\perp}$ is 
\begin{align}
\begin{split}
\label{projection}
    \prod(S_{\bpsi^{*}}(\bO_{i})|\Lambda_{0}^{\perp})
    =&\sum_{k=1}^{K_{i}}\int \left\{1-\Delta\int \lambda_{ik}\left(u | \overline{\bH}_{i, V_{k-1}+w+u}\right) \text{d}u\right\}^{-1}\times\\
    &\Bigg[\Ebb\left\{S_{\bpsi^{*}}(\bO_{i}) \dot{U}_{iu}\left(\bpsi^{*}\right) | \overline{H}_{i, V_{k-1}+w+u}, T_{ik}= u\right\}\\
    &-\Ebb\left\{S_{\bpsi^{*}}(\bO_{i})\dot{U}_{iu}\left(\bpsi^{*}\right) | \overline{H}_{i, V_{k-1}+w+u}, T_{ik}\geqslant u\right\}\Bigg]\times\\
&\left[\textnormal{Var}\left\{U_{i}(\bpsi^{*}) | \overline{H}_{i, V_{k-1}+w+u}, T_{ik}\geqslant u\right\}\right]^{-1}\times\\
&\left[U\left(\bpsi^{*}\right)-\Ebb\left\{U_{i}\left(\bpsi^{*}\right) | \overline{H}_{i, V_{k-1}+w+u}, T_{ik}\geqslant u\right\}\right] \textnormal{d} M_{ik}(u),
\end{split}
\end{align}
where $\dot{U}_{iu}\left(\bpsi^{*}\right)={U}_{i}\left(\bpsi^{*}\right)-\Ebb\{{U}_{i}\left(\bpsi^{*}\right)|\overline{H}_{i, V_{k-1}+w+u}, T_{ik}\geqslant u\}$.\\
To show this, we need to show that the right hand side of equation (\ref{projection}), which is denoted as $A$ is orthogonal to any element in $\Lambda_{0}$. That is equivalent to show that the residual after projecting $S_{\bpsi^{*}}(\bO_{i})$ onto $\Lambda_{0}^{\perp}$ is orthogonal to any element of $\Lambda_{0}^{\perp}$. We denote an arbitrary element of $\Lambda_{0}^{\perp}$ as \begin{align*}
\tilde{G}=\sum_{k=1}^{K_{i}}\int\tilde{c}(\overline{\bH}_{i, V_{k-1}+w+u})\left[U_{i}(\bpsi^{*})-
\Ebb\left\{U_{i}(\bpsi^{*} ) |\overline{\bH}_{i, V_{k-1}+w+u}, V_{i,k}\geqslant V_{i,k-1}+w+u\right\}\right] \text{d}M_{ik}(u),
\end{align*} and show that $\{S_{\bpsi^{*}}(\bO_{i})-A\}\perp \tilde{G}$, that is $\Ebb[(S_{\bpsi^{*}}(\bO_{i})-A) \tilde{G}=0]$. We next show $\Ebb[(S_{\bpsi^{*}}(\bO_{i})\tilde{G}]=\Ebb(A\tilde{G})$ to complete the proof of (\ref{projection}).\\
Firstly,
\begin{align}
\begin{split}
\label{SG}
    &\Ebb[(S_{\bpsi^{*}}(\bO_{i})\tilde{G}]\\
    =& \Ebb \sum_{k=1}^{K_{i}}\int\tilde{c}(\overline{\bH}_{i, V_{k-1}+w+u})S_{\bpsi^{*}}(\bO_{i})\left[U_{i}(\bpsi^{*})-\Ebb\left\{U_{i}(\bpsi^{*} ) |\overline{\bH}_{i, V_{k-1}+w+u}, T_{i,k-1}\geqslant u\right\}\right] \text{d}M_{ik}(u)\\
    =& \Ebb \sum_{k=1}^{K_{i}}\int\tilde{c}(\overline{\bH}_{i, V_{k-1}+w+u})S_{\bpsi^{*}}(\bO_{i})\dot{U}_{iu}\left(\bpsi^{*}\right)\text{d}N_{ik}(u)\\
    -&\Ebb \sum_{k=1}^{K_{i}}\int\tilde{c}(\overline{\bH}_{i, V_{k-1}+w+u})S_{\bpsi^{*}}(\bO_{i})\dot{U}_{iu}\left(\bpsi^{*}\right)\lambda_{ik}\left(u | \overline{\bH}_{i, V_{k-1}+w+u}\right)Y_{ik}(u)
    \text{d}u\\
    =& \Ebb \sum_{k=1}^{K_{i}}\int\tilde{c}(\overline{\bH}_{i, V_{k-1}+w+u}) \Ebb\left\{ S_{\bpsi^{*}}(\bO_{i})\dot{U}_{iu}\left(\bpsi^{*}\right) \mathbbm{1}(u\leqslant T_{i,k-1} \leqslant u+\text{d}u) | \overline{\bH}_{i, V_{k-1}+w+u}, T_{i,k-1}\geqslant u\right\}\\
    -& \Ebb \sum_{k=1}^{K_{i}}\int\tilde{c}(\overline{\bH}_{i, V_{k-1}+w+u}) \Ebb\left\{ S_{\bpsi^{*}}(\bO_{i})\dot{U}_{iu}\left(\bpsi^{*}\right) | \overline{\bH}_{i, V_{k-1}+w+u}, T_{i,k-1}\geqslant u\right\}\lambda_{ik}\left(u | \overline{\bH}_{i, V_{k-1}+w+u}\right)Y_{ik}(u)
    \text{d}u\\
    =& \Ebb \sum_{k=1}^{K_{i}}\int\tilde{c}(\overline{\bH}_{i, V_{k-1}+w+u}) \Ebb\left\{ S_{\bpsi^{*}}(\bO_{i})\dot{U}_{iu}\left(\bpsi^{*}\right)  | \overline{\bH}_{i, V_{k-1}+w+u}, T_{i,k-1}= u\right\}\lambda_{ik}\left(u | \overline{\bH}_{i, V_{k-1}+w+u}\right)Y_{ik}(u)\text{d}u\\
    -& \Ebb \sum_{k=1}^{K_{i}}\int\tilde{c}(\overline{\bH}_{i, V_{k-1}+w+u}) \Ebb\left\{ S_{\bpsi^{*}}(\bO_{i})\dot{U}_{iu}\left(\bpsi^{*}\right) | \overline{\bH}_{i, V_{k-1}+w+u}, T_{i,k-1}\geqslant u\right\}\lambda_{ik}\left(u | \overline{\bH}_{i, V_{k-1}+w+u}\right)Y_{ik}(u)
    \text{d}u\\
    =& \Ebb \sum_{k=1}^{K_{i}}\int\tilde{c}(\overline{\bH}_{i, V_{k-1}+w+u})\Bigg[ \Ebb\left\{ S_{\bpsi^{*}}(\bO_{i})\dot{U}_{iu}\left(\bpsi^{*}\right)  | \overline{\bH}_{i, V_{k-1}+w+u}, T_{i,k-1}= u\right\}\\
    &-\Ebb\left\{ S_{\bpsi^{*}}(\bO_{i})\dot{U}_{iu}\left(\bpsi^{*}\right)  | \overline{\bH}_{i, V_{k-1}+w+u}, T_{i,k-1}\geqslant u\right\}\Bigg]\lambda_{ik}\left(u | \overline{\bH}_{i, V_{k-1}+w+u}\right)Y_{ik}(u)\text{d}u.
\end{split}
\end{align}
Secondly, by Lemma \ref{Lemma1}, we have
\begin{align}
\begin{split}
\label{AG}
    \Ebb(A\tilde{G})
   =&\Ebb\sum_{k=1}^{K_{i}}\int\tilde{c}(\overline{\bH}_{i, V_{k-1}+w+u})\left\{1-\Delta\int \lambda_{ik}\left(u | \overline{\bH}_{i, V_{k-1}+w+u}\right) \text{d}u\right\}^{-1}\\
    &\Bigg[\Ebb\left\{S_{\bpsi^{*}}(\bO_{i}) \dot{U}_{iu}\left(\bpsi^{*}\right) | \overline{H}_{i, V_{k-1}+w+u}, T_{ik}= u\right\}\\
    &-\Ebb\left\{S_{\bpsi^{*}}(\bO_{i})\dot{U}_{iu}\left(\bpsi^{*}\right) | \overline{H}_{i, V_{k-1}+w+u}, T_{ik}\geqslant u\right\}\Bigg]\\
&\times\left[\textnormal{Var}\left\{U_{i}(\bpsi^{*}) | \overline{H}_{i, V_{k-1}+w+u}, T_{ik}\geqslant u\right\}\right]^{-1}\\&\left[U\left(\bpsi^{*}\right)-\Ebb\left\{U_{i}\left(\bpsi^{*}\right) | \overline{H}_{i, V_{k-1}+w+u}, T_{ik}\geqslant u\right\}\right]^{2}\\ &\left\{1-\Delta\int \lambda_{ik}\left(u | \overline{\bH}_{i, V_{k-1}+w+u}\right) \text{d}u\right\}\lambda_{ik}\left(u | \overline{\bH}_{i, V_{k-1}+w+u}\right)Y_{ik}(u)\textnormal{d} u\\
=& \Ebb \sum_{k=1}^{K_{i}}\int\tilde{c}(\overline{\bH}_{i, V_{k-1}+w+u})\Bigg[ \Ebb\left\{ S_{\bpsi^{*}}(\bO_{i})\dot{U}_{iu}\left(\bpsi^{*}\right)  | \overline{\bH}_{i, V_{k-1}+w+u}, T_{i,k-1}= u\right\}\\
    &-\Ebb\left\{ S_{\bpsi^{*}}(\bO_{i})\dot{U}_{iu}\left(\bpsi^{*}\right)  | \overline{\bH}_{i, V_{k-1}+w+u}, T_{i,k-1}\geqslant u\right\}\Bigg]\lambda_{ik}\left(u | \overline{\bH}_{i, V_{k-1}+w+u}\right)Y_{ik}(u)\text{d}u.
\end{split}
\end{align}
By (\ref{SG}) and (\ref{AG}), we have $\forall \tilde{G}, \Ebb[S_{\bpsi^{*}}(\bO_{i})\tilde{G}]=\Ebb(A\tilde{G})$, thus completes the proof of (\ref{projection}). Note that in (\ref{projection}), $S_{\bpsi^{*}}(\bO_{i})$ can be any vector in the Hilbert space, since we have not use the property of score yet. Next, we further simplify $S^{\textnormal{eff}}_{\bpsi^{*}}(\bO_{i})$ by using the facts that
\begin{align*}
\Ebb\left\{\dot{U}_{iu}\left(\bpsi^{*}\right)  | \overline{\bH}_{i, V_{k-1}+w+u}, T_{i,k-1}\geqslant u\right\}=0 \text{  and  }
\Ebb\left\{\dot{U}_{iu}\left(\bpsi^{*}\right)  | \overline{\bH}_{i, V_{k-1}+w+u}, T_{i,k-1}= u\right\}=0.
\end{align*}
Taking the derivative of $\bpsi$ at both sides and using the generalized information equality, we have
\begin{align*}
    \Ebb\left\{S_{\bpsi^{*}}(\bO_{i})\dot{U}_{iu}\left(\bpsi^{*}\right)  | \overline{\bH}_{i, V_{k-1}+w+u}, T_{i,k-1}\geqslant u\right\}+\Ebb\left\{\frac{\partial}{\partial\bpsi} \dot{U}_{iu}\left(\bpsi^{*}\right) | \overline{\bH}_{i, V_{k-1}+w+u}, T_{i,k-1}\geqslant u\right\}=0,
\end{align*}
and
\begin{align*}
    \Ebb\left\{S_{\bpsi^{*}}(\bO_{i})\dot{U}_{iu}\left(\bpsi^{*}\right)  | \overline{\bH}_{i, V_{k-1}+w+u}, T_{i,k-1}= u\right\}+\Ebb\left\{\frac{\partial}{\partial\bpsi} \dot{U}_{iu}\left(\bpsi^{*}\right) | \overline{\bH}_{i, V_{k-1}+w+u}, T_{i,k-1}= u\right\}=0.
\end{align*}
In addition, note that 
\begin{align*}
    &\Ebb\left\{\frac{\partial}{\partial\bpsi} \dot{U}_{iu}\left(\bpsi^{*}\right) | \overline{\bH}_{i, V_{k-1}+w+u}, T_{i,k-1}= u\right\}\\
    =&\Ebb\left(\frac{\partial}{\partial\bpsi}\left[ {U}_{i}\left(\bpsi^{*}\right)-\Ebb\{{U}_{i}\left(\bpsi^{*}\right)|\overline{H}_{i, V_{k-1}+w+u}, T_{ik}\geqslant u\} \right]| \overline{\bH}_{i, V_{k-1}+w+u}, T_{i,k-1}= u\right)\\
    =&0.
\end{align*}
Therefore, by Theorem (\ref{projection}), 
\begin{align*}
    &S^{\textnormal{eff}}_{\bpsi^{*}}(\bO_{i})= \prod(S_{\bpsi^{*}}(\bO_{i})|\Lambda_{0}^{\perp})\\
    =&\sum_{k=1}^{K_{i}}\int \left\{1-\Delta\int \lambda_{ik}\left(u | \overline{\bH}_{i, V_{k-1}+w+u}\right) \text{d}u\right\}^{-1}\times\\
    &\Bigg[\Ebb\left\{S_{\bpsi^{*}}(\bO_{i}) \dot{U}_{iu}\left(\bpsi^{*}\right) | \overline{H}_{i, V_{k-1}+w+u}, T_{ik}= u\right\}
    -\Ebb\left\{S_{\bpsi^{*}}(\bO_{i})\dot{U}_{iu}\left(\bpsi^{*}\right) | \overline{H}_{i, V_{k-1}+w+u}, T_{ik}\geqslant u\right\}\Bigg]\times\\
&\left[\textnormal{Var}\left\{U_{i}(\bpsi^{*}) | \overline{H}_{i, V_{k-1}+w+u}, T_{ik}\geqslant u\right\}\right]^{-1}\times\\
&\left[U\left(\bpsi^{*}\right)-\Ebb\left\{U_{i}\left(\bpsi^{*}\right) | \overline{H}_{i, V_{k-1}+w+u}, T_{ik}\geqslant u\right\}\right] \textnormal{d} M_{ik}(u)\\
=&-\sum_{k=1}^{K_{i}}\int \left\{1-\Delta\int \lambda_{ik}\left(u | \overline{\bH}_{i, V_{k-1}+w+u}\right) \text{d}u\right\}^{-1}\times\\
    &\Bigg[\Ebb\left\{\frac{\partial}{\partial\bpsi} \dot{U}_{iu}\left(\bpsi^{*}\right) | \overline{H}_{i, V_{k-1}+w+u}, T_{ik}= u\right\}
    -\Ebb\left\{\frac{\partial}{\partial\bpsi} \dot{U}_{iu}\left(\bpsi^{*}\right) | \overline{H}_{i, V_{k-1}+w+u}, T_{ik}\geqslant u\right\}\Bigg]\times\\
&\left[\textnormal{Var}\left\{U_{i}(\bpsi^{*}) | \overline{H}_{i, V_{k-1}+w+u}, T_{ik}\geqslant u\right\}\right]^{-1}\times\\
&\left[U\left(\bpsi^{*}\right)-\Ebb\left\{U_{i}\left(\bpsi^{*}\right) | \overline{H}_{i, V_{k-1}+w+u}, T_{ik}\geqslant u\right\}\right] \textnormal{d} M_{ik}(u)\\
=&-\sum_{k=1}^{K_{i}}\int \left\{1-\Delta\int \lambda_{ik}\left(u | \overline{\bH}_{i, V_{k-1}+w+u}\right) \text{d}u\right\}^{-1}\times\\
    &\Ebb\left\{\frac{\partial}{\partial\bpsi} \dot{U}_{iu}\left(\bpsi^{*}\right) | \overline{H}_{i, V_{k-1}+w+u}, T_{ik}= u\right\}\times
\left[\textnormal{Var}\left\{U_{i}(\bpsi^{*}) | \overline{H}_{i, V_{k-1}+w+u}, T_{ik}\geqslant u\right\}\right]^{-1}\times\\
&\left[U\left(\bpsi^{*}\right)-\Ebb\left\{U_{i}\left(\bpsi^{*}\right) | \overline{H}_{i, V_{k-1}+w+u}, T_{ik}\geqslant u\right\}\right] \textnormal{d} M_{ik}(u),
\end{align*}
and that completes the proof of Theorem \ref{efficient score}. Compared with (\ref{estimating equation}), we obtain the optimal choice of $c(\overline{\bH}_{i, V_{k-1}+w+u})$, as is shown in (\ref{copt}).
\section{Proof of Theorem \ref{DR}}
\label{DR_proof}
To show the double-robustness property of the proposed estimator, we need to show $$\Ebb\{G(\bpsi^{*},\bO_{i})\}=0$$ when either $\lambda_{ik}\left(u | \overline{\bH}_{i, V_{k-1}+w+u}\right)$ or $\Ebb\left\{U_{i}\left(\bpsi^{*}\right) | \overline{H}_{i, V_{k-1}+w+u}, T_{ik}\geqslant u\right\}$ is correctly specified.\\
Firstly, when $\lambda_{ik}\left(u | \overline{\bH}_{i, V_{k-1}+w+u}\right)$ is correctly specified, $M_{ik}(u)$ is a zero mean martingale with respect to the filtration $\mathscr{F}_{i,V_{k-1}+w+u}$. According to Lemma \ref{Lemma1}, because $$c(\overline{\bH}_{i, V_{k-1}+w+u})\left[U_{i}(\bpsi)-\Ebb\left\{U_{i}(\bpsi) |\overline{\bH}_{i, V_{k-1}+w+u}, T_{ik}\geqslant u\right\}\right]$$ is bounded and $\mathscr{F}_{i,V_{k-1}+w+u}$-predictable (even if $\Ebb\left\{U_{i}\left(\bpsi^{*}\right) | \overline{H}_{i, V_{k-1}+w+u}, T_{ik}\geqslant u\right\}$ is incorrectly specified), then $G(\bpsi^{*},\bO_{i})$ is also a zero mean martingale.\\
Secondly, when $\Ebb\left\{U_{i}\left(\bpsi^{*}\right) | \overline{H}_{i, V_{k-1}+w+u}, T_{ik}\geqslant u\right\}$ is correctly specified, and let $\lambda_{ik}^{*}\left(u | \overline{\bH}_{i, V_{k-1}+w+u}\right)$ be the incorrectly specified model for $\lambda_{ik}\left(u | \overline{\bH}_{i, V_{k-1}+w+u}\right)$. We have
\begin{align*}
    &\Ebb\{G(\bpsi^{*},\bO_{i})\}\\
    =&\Ebb\sum_{k=1}^{K_{1}} \int c(\overline{\bH}_{i, V_{k-1}+w+u})\left[U_{i}(\bpsi)-E\left\{U_{i}(\bpsi) |\overline{\bH}_{i, V_{k-1}+w+u}, T_{ik}\geqslant u\right\}\right]\\
    &\{\text{d}N_{ik}(u)-\lambda_{ik}^{*}\left(u | \overline{\bH}_{i, V_{k-1}+w+u}\right)Y_{ik}(u)\text{d}u\}\\
    =&\Ebb\sum_{k=1}^{K_{1}} \int c(\overline{\bH}_{i, V_{k-1}+w+u})\left[U_{i}(\bpsi)-E\left\{U_{i}(\bpsi) |\overline{\bH}_{i, V_{k-1}+w+u}, T_{ik}\geqslant u\right\}\right]\\
    & \{\text{d}N_{ik}(u)-\lambda_{ik}\left(u | \overline{\bH}_{i, V_{k-1}+w+u}\right)Y_{ik}(u)\text{d}u\}\\
    +&\Ebb\sum_{k=1}^{K_{1}} \int c(\overline{\bH}_{i, V_{k-1}+w+u})\left[U_{i}(\bpsi)-E\left\{U_{i}(\bpsi) |\overline{\bH}_{i, V_{k-1}+w+u}, T_{ik}\geqslant u\right\}\right]\times\\
    &\{\lambda_{ik}\left(u | \overline{\bH}_{i, V_{k-1}+w+u}\right)-\lambda_{ik}^{*}\left(u | \overline{\bH}_{i, V_{k-1}+w+u}\right)Y_{ik}(u)\text{d}u\}\\
    =& 0+\Ebb\sum_{k=1}^{K_{1}} \int c(\overline{\bH}_{i, V_{k-1}+w+u})\Ebb\left(\left[U_{i}(\bpsi)-E\left\{U_{i}(\bpsi) |\overline{\bH}_{i, V_{k-1}+w+u}, T_{ik}\geqslant u\right\}\right]|\overline{\bH}_{i, V_{k-1}+w+u}, T_{ik}\geqslant u\right)\\
    &\times\{\lambda_{ik}\left(u | \overline{\bH}_{i, V_{k-1}+w+u}\right)-\lambda_{ik}^{*}\left(u | \overline{\bH}_{i, V_{k-1}+w+u}\right)Y_{ik}(u)\text{d}u\}\\
    =&0.
\end{align*}
In addition, the locally efficient estimator is obtained by solving the parameter of interest from the equation that setting the empirical mean of efficient score to 0, which is exactly the estimating equation when we replace $c(\overline{\bH}_{i, V_{k-1}+w+u})$ by $c^{\text{opt}}(\overline{\bH}_{i, V_{k-1}+w+u})$ in (\ref{replaced estimating equation}). Therefore, our proposed estimator is locally efficient. By Theorem 4.1 of \citep{tsiatis2007semiparametric}, the semiparametric efficiency bound is given by $[\Ebb\{S^{\textnormal{eff}}_{\bpsi^{*}}(\bO_{i})S^{\textnormal{eff}\intercal}_{\bpsi^{*}}(\bO_{i})\}]^{-1}$.

\section{Proof of Theorem \ref{Theorem3}}
\label{theorem3}
First, we need to show it is a consistent estimator. Towards this end, we show 
\begin{align*}
    \Ebb\left\{\frac{\Delta_{i}}{S_{Ci}(\tau_{i}|\overline{\bH}_{\tau_{i}})}G(\bpsi,\bO_{i})\right\}=0.
\end{align*}
By applying the Law of total expectation, we have
\begin{align*}
    \Ebb\left\{\frac{\Delta_{i}}{S_{Ci}(\tau_{i}|\overline{\bH}_{\tau_{i}})}G(\bpsi,\bO_{i})\right\}
    =&\Ebb\left[\left\{\frac{\Delta_{i}}{S_{Ci}(\tau_{i}|\overline{\bH}_{\tau_{i}})}G(\bpsi,\bO_{i})|\bO_{i}\right\}\right]\\
    =&\Ebb\left\{\frac{\Ebb\{\Delta_{i}|\bO_{i}\}}{S_{Ci}(\tau_{i}|\overline{\bH}_{\tau_{i}})}G(\bpsi,\bO_{i})\right\}\\
    =&\Ebb\left\{1\times G(\bpsi,\bO_{i})\right\}=0.
\end{align*}
Then, as shown in Supplementary material \ref{DR_proof}, when $h\left\{u,U_{i}(\bpsi^{*}), \overline{\bH}_{i, V_{k-1}+w+u}\right\}=c(\overline{\bH}_{i, V_{k-1}+w+u})\\U_{i}(\bpsi^{*})$, and given $\lambda_{iC}\left(u | \overline{\bH}_{iu}\right)$ is correctly specified, the estimator is doubly-robust.
\end{appendices}
\bibliographystyle{apalike}
\bibliography{ref}
\end{document}